\documentclass[reprint,prb,superscriptaddress,showpacs]{revtex4-2}%
\usepackage{graphicx}
\usepackage{color}
\usepackage{amsmath}
\usepackage{amsfonts}
\usepackage{amssymb}
\usepackage{framed}
\usepackage{placeins}
\usepackage{float}
\usepackage[normalem]{ulem}
\providecommand{\U}[1]{\protect\rule{.1in}{.1in}}
\makeatletter
\renewcommand*{\fnum@figure}{{\normalfont\bfseries \figurename~\thefigure}}
\renewcommand*{\@caption@fignum@sep}{\textbf{ : }}
\makeatother
\usepackage{hyperref}
\hypersetup{
    colorlinks=true,
    linkcolor=blue,
    filecolor=magenta,      
    urlcolor=cyan,
}
\begin{document}

\title{Discovery of Kondo-Kramers Nodal Line Topological State in strongly correlated material CeAuAl$_3$}

\author{Hari Bhandari}
\affiliation{Department of Physics and Astronomy, Rice University, Houston, Texas 77005, USA}
\affiliation{Rice Center for Quantum Materials, Rice University, Houston, Texas 77005, USA}
\affiliation{Rice Laboratory for Emergent Magnetic Materials and Smalley-Curl Institute, Rice University, Houston, Texas 77005, USA}

\author{Kevin Allen}
\affiliation{Department of Physics and Astronomy, Rice University, Houston, Texas 77005, USA}
\affiliation{Rice Center for Quantum Materials, Rice University, Houston, Texas 77005, USA}
\affiliation{Rice Laboratory for Emergent Magnetic Materials and Smalley-Curl Institute, Rice University, Houston, Texas 77005, USA}

\author{Kuan-Sen Lin}
\affiliation{Department of Physics and Astronomy, Rice University, Houston, Texas 77005, USA}
\affiliation{Rice Center for Quantum Materials, Rice University, Houston, Texas 77005, USA}
\affiliation{Rice Laboratory for Emergent Magnetic Materials and Smalley-Curl Institute, Rice University, Houston, Texas 77005, USA}
\affiliation{Extreme Quantum Materials Alliance, Rice University, Houston, Texas 77005, USA}

\author{Yuan Fang}
\affiliation{Department of Physics and Astronomy, Rice University, Houston, Texas 77005, USA}
\affiliation{Rice Center for Quantum Materials, Rice University, Houston, Texas 77005, USA}
\affiliation{Rice Laboratory for Emergent Magnetic Materials and Smalley-Curl Institute, Rice University, Houston, Texas 77005, USA}
\affiliation{Extreme Quantum Materials Alliance, Rice University, Houston, Texas 77005, USA}

\author{ Arushi}
\affiliation{Department of Physics and Astronomy, Rice University, Houston, Texas 77005, USA}
\affiliation{Rice Center for Quantum Materials, Rice University, Houston, Texas 77005, USA}
\affiliation{Rice Laboratory for Emergent Magnetic Materials and Smalley-Curl Institute, Rice University, Houston, Texas 77005, USA}

\author{Karthik Rao}
\affiliation{Department of Physics and Astronomy, Rice University, Houston, Texas 77005, USA}
\affiliation{Rice Center for Quantum Materials, Rice University, Houston, Texas 77005, USA}
\affiliation{Rice Laboratory for Emergent Magnetic Materials and Smalley-Curl Institute, Rice University, Houston, Texas 77005, USA}

\author{Mounica Mahankali}
\affiliation{Department of Physics and Astronomy, Rice University, Houston, Texas 77005, USA}
\affiliation{Rice Center for Quantum Materials, Rice University, Houston, Texas 77005, USA}
\affiliation{Rice Laboratory for Emergent Magnetic Materials and Smalley-Curl Institute, Rice University, Houston, Texas 77005, USA}
\affiliation{Extreme Quantum Materials Alliance, Rice University, Houston, Texas 77005, USA}

\author{Sanu Mishra}
\affiliation{Department of Physics and Astronomy, Rice University, Houston, Texas 77005, USA}
\affiliation{Rice Center for Quantum Materials, Rice University, Houston, Texas 77005, USA}
\affiliation{Rice Laboratory for Emergent Magnetic Materials and Smalley-Curl Institute, Rice University, Houston, Texas 77005, USA}

\author{Jian-Xin Zhu}
\affiliation{Theoretical Division, Los Alamos National Laboratory, NM, 87545, USA}

\author{Christopher Lane}
\affiliation{Theoretical Division, Los Alamos National Laboratory, NM, 87545, USA}

\author{Qimiao Si}
\affiliation{Department of Physics and Astronomy, Rice University, Houston, Texas 77005, USA}
\affiliation{Rice Center for Quantum Materials, Rice University, Houston, Texas 77005, USA}
\affiliation{Rice Laboratory for Emergent Magnetic Materials and Smalley-Curl Institute, Rice University, Houston, Texas 77005, USA}
\affiliation{Extreme Quantum Materials Alliance, Rice University, Houston, Texas 77005, USA}

\author{Emilia Morosan}
\email{Corresponding author: emorosan@rice.edu}
\affiliation{Department of Physics and Astronomy, Rice University, Houston, Texas 77005, USA}
\affiliation{Rice Center for Quantum Materials, Rice University, Houston, Texas 77005, USA}
\affiliation{Rice Laboratory for Emergent Magnetic Materials and Smalley-Curl Institute, Rice University, Houston, Texas 77005, USA}

\date{\today}
\begin{abstract}

Kramers nodal lines (KNLs) are predicted in all achiral, non-centrosymmetric crystal structures, and yet there are very few experimental realizations of KNLs, only in weakly correlated systems. Especially, their realization in strongly correlated systems remains unexplored. Here we investigate KNL topology in the non-centrosymmetric, achiral Kondo lattice CeAuAl$_3$, using magnetotransport, thermodynamic measurements, density functional theory (DFT), and tight-binding toy model calculations. We find that CeAuAl$_3$ hosts KNLs in the $\mathit{\Gamma}\text{-}Z$ and $X\text{-}P$ directions, protected by time-reversal and mirror symmetries. Breaking of these crystalline symmetries, along with time-reversal symmetry via spin polarization, lifts the KNL degeneracy, which, in turn, generates Berry curvature resulting in an intrinsic anomalous Hall effect (AHE). More importantly, CeAuAl$_3$ shows large AHE not only below the magnetic ordering temperature $T_N$ but rather below the Kondo temperature. This underscores the role of the Kondo correlations in bringing the KNLs at the Fermi surface rather than at accidental energies, as would be expected without correlations. Our work identifies a new topological state, Kondo-Kramers Nodal Line, with CeAuAl$_3$ being the first compound explicitly identified to host this new correlated topological state, and demonstrating the expected AHE associated with it. More generally, our work outlines the pathway for finding  Kondo-KNL systems in all compounds that combine achiral, non-centrosymmetric structure with Kondo correlations.

\end{abstract}

\maketitle

\section{Introduction}\label{sec:1}

Strongly correlated systems give rise to a wide variety of emergent quantum phenomena \cite{paschen2021quantum,checkelsky2024flat}. Heavy fermion systems are of particular interest because the interplay between \textit{f} electrons and itinerant conduction electrons can give rise to unconventional quantum criticality \cite{gegenwart2008quantum,coleman2005quantum}, and, when combined with non-trivial topology \cite{siddiquee2023breakdown,guo2018evidence,cheng2023observation,checkelsky2024flat}, results in a new class of topological matter such as Weyl–Kondo semimetals \cite{dzsaber2017kondo,lai2018weyl,dzsaber2021giant,grefe2020weyl}.

Recently, a distinct class of topological band crossings, Kramers nodal lines (KNLs), has been proposed to exist in all achiral, non-centrosymmetric materials with spin-orbit coupling (SOC) \cite{xie2021kramers}. These systems host doubly-degenerate KNLs that connect time reversal invariant momentum (TRIM) points and are protected by achiral symmetries (mirror or roto-inversion symmetry) and time-reversal symmetry. Although the KNL topology has been studied in weakly interacting systems \cite{kurumaji2025electronic,zhang2023kramers,zhang2025kramers,sarkar2023charge,shang2022unconventional,sarkar2026kramers}, the interplay between Kondo correlations and KNL topology remains entirely unexplored. Establishing this interplay is essential for clarifying how electron correlations and crystalline symmetry jointly dictate the emergence of new topological phases \cite{chen2022topological}. 

CeAuAl$_3$ is an ideal platform to study KNL topology within a strongly correlated electron system. Due to its achiral, non-centrosymmetric crystal structure (tetragonal space group \#107 \textit{I}4\textit{mm}), it hosts KNLs \cite{xie2021kramers, hirschmann2021symmetry}. The nodal degeneracies are pinned at TRIMs, and some are extended along high symmetry lines while others can move freely on the mirror-invariant plane. Importantly, strong electronic correlations arising from the Kondo effect renormalize the electronic bands in the vicinity of Fermi energy $E_{\mathrm{F}}$, pinning the heavy \textit{f}-derived KNLs close to $E_{\mathrm{F}}$, where they can directly dictate the low-energy transport. While previous heavy-fermion studies reported Weyl-Kondo points or general nodal lines \cite{lai2018weyl,dzsaber2021giant,shen2026giant,cao2020trivial}, this study represents the first explicit identification of KNL topology in a Kondo lattice system and identifies the new correlated topological state Kramers-Kondo nodal line state.

In this work, we use anomalous Hall conductivity (AHC) measurements for CeAuAl$_3$ together with DFT and tight-binding toy models calculations to demonstrate the interplay between KNL and Kondo correlations that result in both extrinsic and intrinsic AHC contributions. At the lowest measured temperature, the AHC reaches 160 S/cm, a value comparable to those observed in several heavy fermion ferromagnets \cite{li2025extrinsic,cheng2023observation,xu2024tunable}. This large AHC in a Kondo antiferromagnet is driven primarily by a dominant intrinsic Berry curvature that emerges below the Kondo coherence temperature. Our combined DFT and model calculations support the transport measurements and indicate that the observed AHC originates from KNL topology. These results offer a novel pathway for investigating the interplay between topology and magnetism in antiferromagnetic (AFM) Kondo lattices. This work represents the first systematic demonstration of topological KNLs crossings associated with the intrinsic anomalous Hall effect in a strongly correlated system, positioning CeAuAl$_3$ as a prototypical KNL Kondo system.

\begin{figure*}[!ht]
\centering
\includegraphics[width=0.8\textwidth]{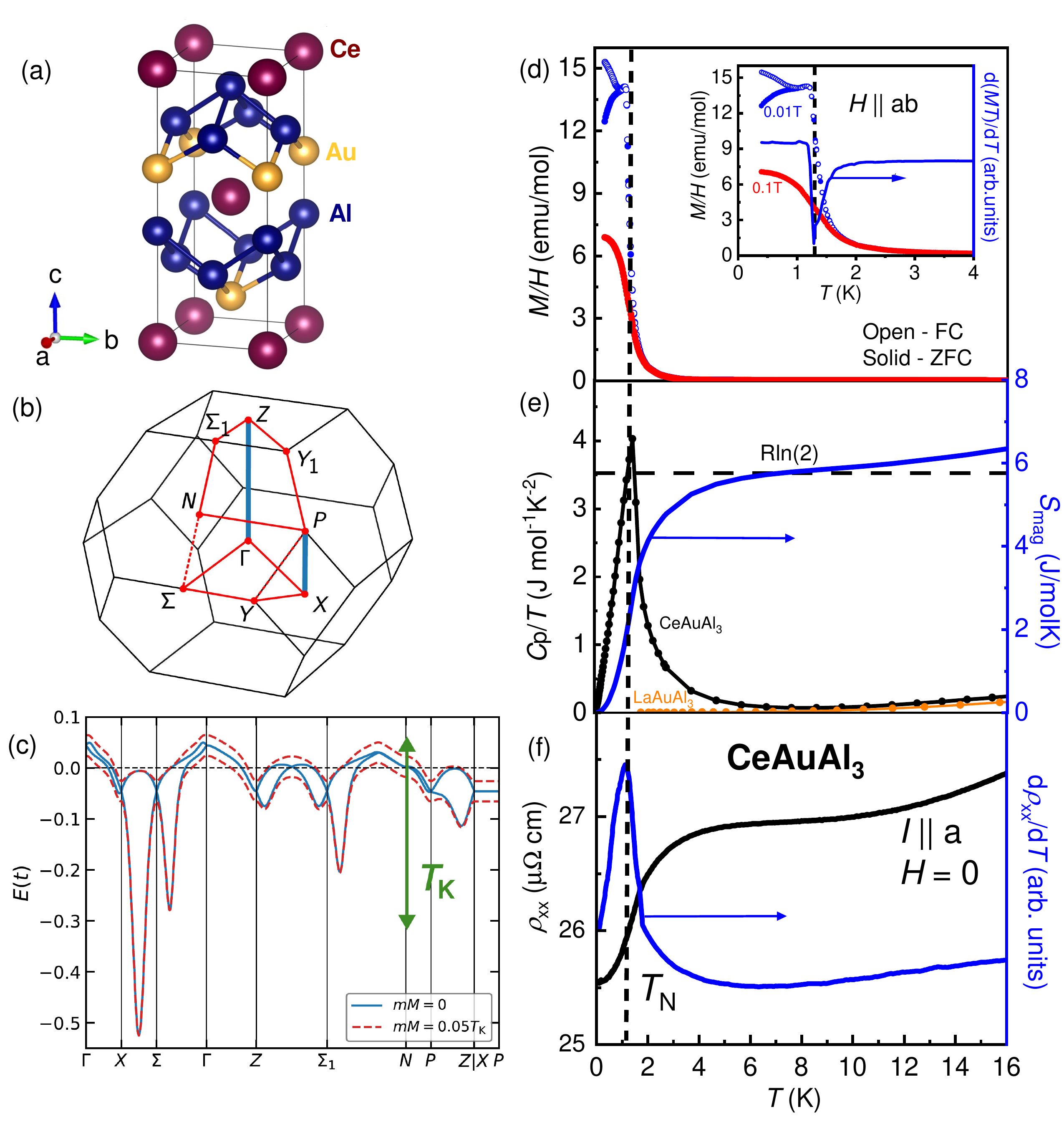}
\caption{\small \textbf{Physical properties and toy model calculations for CeAuAl$_3$.}
(a) Crystal structure of CeAuAl$_3$ (b) 3D Brioullion Zone with high symmetry lines highlighted in red, where blue lines correspond to KNLs along high symmetry lines. (c) Saddle point solutions with effective Zeeman strength $mM=0$ and $mM=0.05~ T_{\mathrm{K}}$. The KNLs along $\mathit{\Gamma}\text{-}Z$ and $X-P$ are gapped when $mM \neq 0$ (dashed lines).(d) Zero-field-cooled (ZFC) and field-cooled (FC) magnetic susceptibility $(M/H)$ for $T$ =  0.4 - 16 K and $H$ = 0.01 T (blue) and 0.1 T (red), \textit{H} $||$ \textit{ab}. Inset: low temperature $(M/H)$ (left axis) and $\mathrm{d}(MT)/\mathrm{d}T$ (right axis). (e) $H$ = 0 specific heat scaled by temperature $C_{p}/T$ (left axis) for CeAuAl$_3$ (black) and LaAuAl$_3$ (orange) and magnetic entropy S$_{mag}$ (right axis). (f) $H$ = 0 electrical resistivity $\rho(T)$ (left axis) and $\mathrm{d}\rho_{xx}/\mathrm{d}T$ (right panel). Vertical dashed line indicates $T_N$ = 1.3 K.}
\label{Fig1}
\end{figure*}

\section{Results}\label{S3}

While the physical properties of CeAuAl$_3$ have been widely investigated \cite{klicpera2014study,aoki2000anomalous,vcermak2019magnetoelastic,adroja2015muon,sugawara1999magnetic,ullah2024electronic,paschen1998transport,vonlanthen199927al,deyerling2023electronic,franz2016single,mock1999low,liu2018lattice}, the unique band topology of the KNLs and its direct connection to the electronic properties have not been explored. More importantly, experimental realization of KNL topology in a Kondo system is currently lacking. CeAuAl$_3$ is the first system in which signatures of both Kondo behavior and KNL topology are evident, as illustrated in Fig. \ref{Fig1}. CeAuAl$_3$ crystallizes in the non-centrosymmetric, tetragonal BaNiSn$_3$-type structure (Fig. \ref{Fig1}a), where Ce moments couple ferromagnetically within the \textit{ab} plane and form an incommensurate spiral structure along the \textit{c} axis, with the propagation vector \textbf{k} = (0,0,0.52(1)) \cite{adroja2015muon}. The most notable consequence of the crystal structure is illustrated by our toy-model calculations (Fig. \ref{Fig1}c) for the paramagnetic (PM, \textit{mM} = 0) and spin-polarized (SP, $mM=0.05 T_{\mathrm{K}}$ states), where $mM$ is the effective Zeeman field. The 3D Brillouin Zone with high symmetry lines highlighted in red is shown in Fig. \ref{Fig1}b. Using the Bilbao Crystallographic Server \cite{elcoro2017double}, we verify that, in Cartesian coordinates, along $(0, 0, k_z)$ and $\left(\frac{2\pi}{2a}, \frac{2\pi}{2a}, k_z\right)$, the energy bands must be twofold degenerate, consistent with previous studies~\cite{xie2021kramers, hirschmann2021symmetry}. In addition, both $(0, 0, k_z)$ and $\left(\frac{2\pi}{2a}, \frac{2\pi}{2a}, k_z\right)$ pass through TRIMs, directly linking the twofold degeneracies to KNLs. We have verified this through our model calculations (Fig. \ref{Fig1}c), in which the $\mathit{\Gamma}\text{-}Z$ and $X\text{-}P$ lines have twofold degenerate nodal lines when $mM=0$ (solid blue lines). We note that $(0, 0, k_z)$ includes $\mathit{\Gamma}\text{-}Z$ and $\left(\frac{2\pi}{2a}, \frac{2\pi}{2a}, k_z\right)$ includes $X\text{-}P$. In addition to $\mathit{\Gamma}\text{-}Z$ and $X\text{-}P$ that are rotation invariant in reciprocal space, there are also KNLs that are not pinned along rotation invariant lines. A mirror invariant plane in reciprocal space that hosts such KNLs is the plane that contains $\mathit{\Gamma}$, $Z$, and $N$. In the presence of an external magnetic field $mM=0.05~ T_{\mathrm{K}}$ (Fig. \ref{Fig1}c), the time-reversal symmetry breaks, causing the KNLs to split. Due to the strongly renormalized energy scale of the heavy fermion bands, a small Zeeman term \textit{mM}, representing an external field, gaps out the KNLs.  Consequently, the gapped KNLs act as sources of Berry curvature pinned close to the Fermi energy by the Kondo effect, as shown in Fig. \ref{Fig1}c (dashed red lines). The model parameters are chosen to qualitatively mimic the low-energy energetics of the DFT band structure (discussed in a later section). In particular, the toy model captures key features: (i)~a gap around the Fermi energy; (ii) hole-like dispersions of both conduction and valence bands near $\mathit{\Gamma}$; and (iii)~low-energy conduction-band states along $X\text{-}P$. Overall, the toy model provides a qualitative description of the key low-energy features of the DFT band structure. The details of the toy model and parton calculations are further discussed in Supplementary Materials S6.  

An important feature of the band structure calculation is the pinning of the KNLs close to the Fermi surface due to Kondo effect. This is evidenced in our thermodynamic and transport measurements shown in Fig. \ref{Fig1}d-f. Magnetic susceptibility (Fig. \ref{Fig1}d) shows magnetic order below $T_\mathrm{N}$ = 1.3~\,K, confirmed by the \textit{H} = 0 specific heat data (black, Fig. \ref{Fig1}e) and resistivity (black, Fig. \ref{Fig1}f). The antiferromagnetic order below $T_\mathrm{N}$  = 1.3~\,K, consistent with previous reports \cite{paschen1998transport,adroja2015muon,franz2016single,aoki2000anomalous}, is more clearly evidenced by sharp peaks in the derivatives d\textit{(MT)}/d\textit{T} (inset, Fig. \ref{Fig1}d) and d$\rho$/d\textit{T} (blue, Fig. \ref{Fig1}f). Below ${T_N}$, the zero-field-cooled (ZFC) and field-cooled (FC) susceptibility bifurcates, indicating the presence of a small ferromagnetic (FM) component arising from an incommensurate spiral ground state magnetic structure \cite{adroja2015muon}. 

\begin{figure*}[!ht]
\centering
\includegraphics[width=\textwidth]{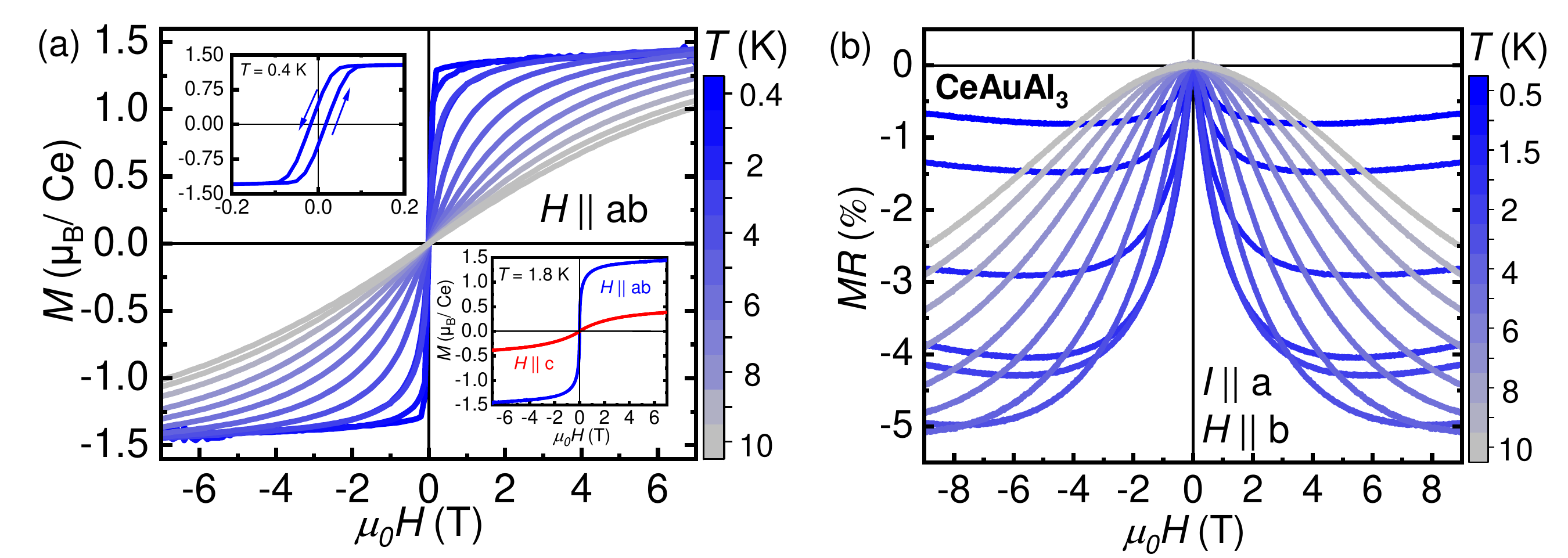}
\caption{\small \textbf{Magnetization and magnetoresistance of CeAuAl$_3$.} (a) Magnetization as a function of magnetic field for $H \parallel ab$ measured from 0.4--10~K. Upper inset shows the hysteresis loop observed at \textit{T}= 0.4 K. Lower inset shows the magnetization at \textit{T} = 1.8 K for \textit{H} $||$ \textit{ab} and \textit{H} $||$ \textit{c}. (b)  Magnetoresistance as a function of magnetic field for \textit{I} $||$ \textit{a} and \textit{H} $||$ \textit{b} measured from $T$ = 0.5-10 K.}

\label{Fig2}
\end{figure*}

A comparison of the specific heat \textit{C$_p$} for both CeAuAl$_3$ (black) and the non-magnetic analog LaAuAl$_3$ (orange) points to signatures of strong electronic correlations, evident from an enhanced Sommerfeld coefficient $\gamma = 73\text{ mJ mol}^{-1}\text{K}^{-2}$ (Supplementary Materials Fig.~S3a) and from the magnetic entropy $S_m$ at $T_\mathrm{N}$  (right axis, Fig.~\ref{Fig1}e). $S_{mag}$ (blue), calculated from the magnetic contribution to the CeAuAl$_3$ specific heat after subtracting the phonon contribution from the non-magnetic analouge LaAuAl$_3$ (orange), reaches only $0.4\,R \ln 2$ at $T_\mathrm{N}$ , while the full doublet entropy $R \ln 2$ is recovered by \textit{T} = 7 K. This behavior indicates a significant Kondo screening of the Ce moments and is consistent with the reduced $M_{sat}$ value observed in $M(H)$ measurements (Fig.~ \ref{Fig2}a). The Kondo temperature $T_{\text{K}} \approx 3~\text{K}$ is estimated from the relation $S_{\text{mag}}(0.5~ T_{\text{K}}) = \frac{1}{2} R \ln 2$ \cite{desgranges1982specific}.

The resistivity shows a broad shoulder around \textit{T} = 3.5 K (black Fig.~\ref{Fig1}f), commonly observed in heavy fermion compounds at the onset of Kondo correlations \cite{adroja2015muon,paschen1998transport,yang2008scaling,allen2026chiral}. At lower temperatures, the resistivity decrease is due to coherent scattering and the loss of spin-disorder scattering as the system approaches magnetic order. The Kondo correlations are further supported by the behavior of the resistivity in applied magnetic field and comparison with the non-magnetic analogue LaAuAl$_3$ (Supplementary Materials S2 and Supplementary Materials Fig.~S2a-b).

The field-dependent behavior of the magnetization \textit{M(H)} and magnetoresistance \textit{MR} of CeAuAl$_3$ provide further insight into the signatures of the Kondo effect (Fig. \ref{Fig2}). The $M(H)$ isotherms (Fig. \ref{Fig2}a) demonstrate the crystal electric field (CEF) anisotropy.  For \textit{H} $||$ \textit{ab}, $M(H)$ increases sharply at low fields and approaches the spin-polarized (SP) state with a small field $H~>$ 0.1 T. A small hysteresis for $M(H\parallel ab)$ (top inset) and $T~<~T_N$ is due to the small ferromagnetic component at low temperatures, as was also indicated by the ZFC-FC susceptibility irreversibility (Fig. \ref{Fig1}d). Kondo screening and the CEF effects result in the SP magnetization saturating at $1.42~\,\mu_{\mathrm{B}}/\mathrm{Ce}$, consistent with previous studies \cite{adroja2015muon}, less than the theoretical value $\mu_{\mathrm{sat}}(\mathrm{Ce}^{3+}) = 2.14 \, \mu_{\mathrm{B}}$ \cite{adroja2004reduced,ueland2018reduction,kratochvilova2015coexistence}. A comparison of the \textit{T} = 1.8 K $M(H)$ data for $H \parallel ab$ (blue) and $H \parallel c$ (red) in the bottom inset shows that the in-plane magnetization is more than three times larger than \textit{M}($H\parallel c$), which, together with the susceptibility measurements (Fig. \ref{Fig1}d), indicates easy-plane anisotropy in CeAuAl$_3$.


Further evidence of correlations is provided by the \textit{MR} measurements shown in Fig. \ref{Fig2}b (for current $I \parallel a$ and field $H \parallel b$) and Supplementary Materials Fig. S4b (for current $I \parallel a$ and field $H \parallel c$). \textit{MR} is negative for all temperatures between 0.5 and 10 K. In heavy fermion compounds \cite{sluchanko2015anomalies}, negative \textit{MR} is commonly related to the suppression of spin-flip scattering of conduction electrons \cite{sluchanko2015anomalies} and the Kondo effect \cite{kondo1964resistance}. The negative \textit{MR} increases with cooling and reaches -5\%  at \textit{T} = 4 K, close to the Kondo coherence temperature, due to the crossover from incoherent single-impurity scattering to a coherent Kondo lattice. Below the Kondo coherence temperature, the negative \textit{MR} shows a weak upturn at higher fields.


\begin{figure*}[!t]
\centering
\includegraphics[width=1\textwidth]{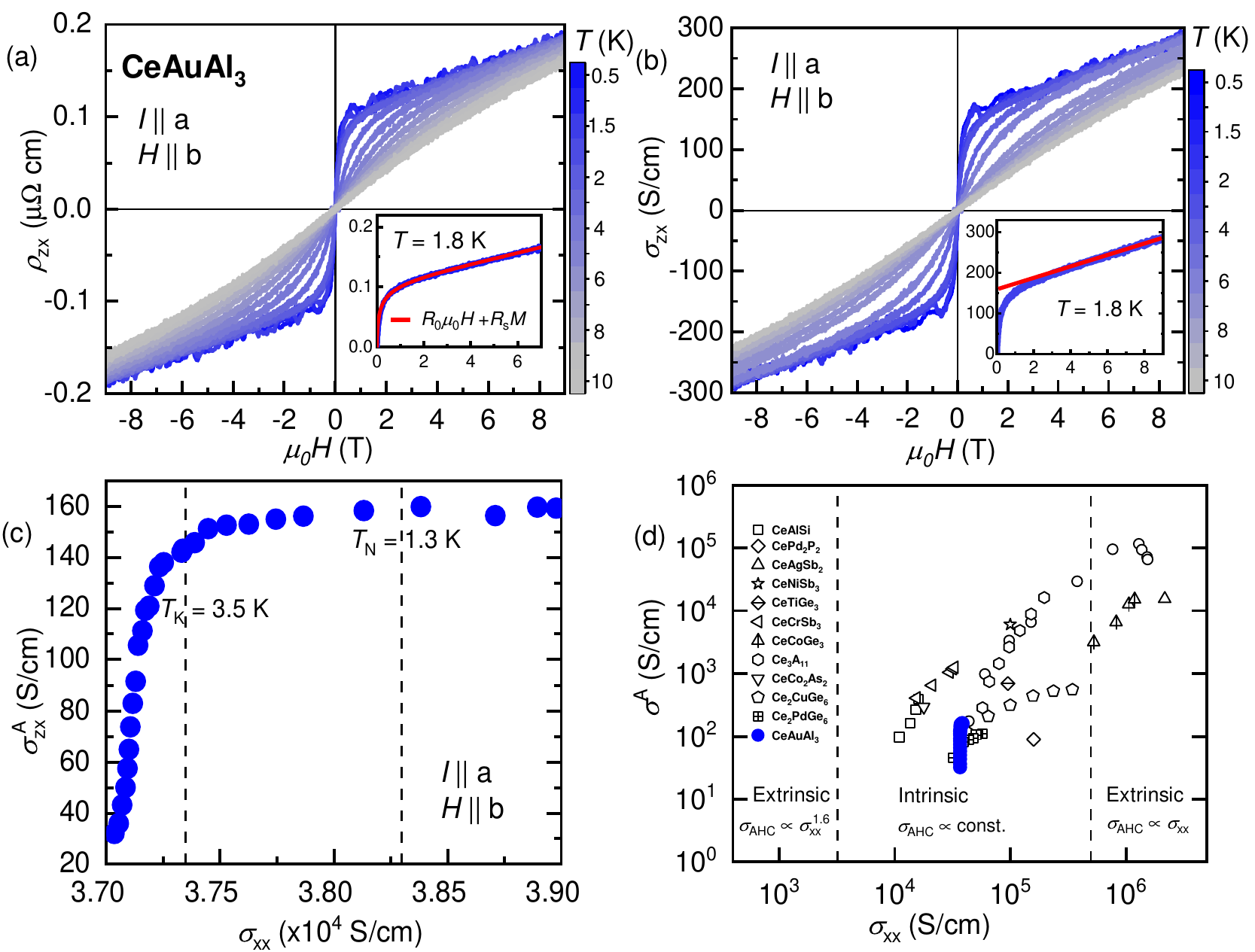}
\caption{\small \textbf{Hall resistivity and Hall conductivity of CeAuAl$_3$.}
(a) Hall resistivity ($\rho_{zx}$) as a function of magnetic field for \textit{H} $||$ \textit{b} and \textit{I} $||$ \textit{a} at selected temperatures between 0.5 K to 10 K. 
(b) Hall conductivity ($\sigma_{zx}$) as a function of magnetic field  for \textit{H} $||$ \textit{b} and \textit{I} $||$ \textit{a} at selected temperatures between 0.5 K and 10 K. (c) Anomalous Hall conductivity ($\sigma^A_{zx}$) as a function of longitudinal conductivity $\sigma_{xx}$ (values taken from $H$ = 0) for \textit{H} $||$ \textit{b} and \textit{I} $||$ \textit{a} for selected temperatures between 0.5 K and 10 K. The Kondo temperature ($T_K$ = 3.5 K) was estimated from resistivity and specific heat measurements. (d) Anomalous Hall conductivity $\sigma^{A}$  \textit{vs} $\sigma_{xx}$ for CeAuAl$_3$ (full blue circles) compared alongside reported Ce-based compounds \cite{li2025extrinsic,shen2026giant, cheng2023observation,li2026giant, kotegawa2024large,alam2023sign,kumar2021giant,allen2026large}.}
\label{Fig3}
\end{figure*}

Building on our thermodynamic and \textit{MR} measurements showing signatures of \textit{c-f} hybridization, we performed Hall effect measurements (Fig. \ref{Fig3}) to explain the non-trivial topological features observed in the electronic structure. The total Hall resistivity $\rho{(H)}$ of a magnetic material is described by \cite{nagaosa2010anomalous}:
\begin{equation}
\rho(H) = \rho^{\text{OHE}}(H) + \rho^{\text{AHE}}(M) = R_0\mu_0H + R_S  M
\end{equation}
where $R_0\mu_0H$ and $R_S  M$ represent the ordinary Hall effect (OHE) and anomalous Hall effect (AHE) contributions, with $R_0$ and $R_S$ denoting the ordinary and anomalous Hall coefficients, respectively.

The easy-plane Hall resistivity $\rho_{zx}$ ($H~||~b$, $I~||~a$) (Fig.~\ref{Fig3}a) follows the magnetization profile (Fig.~\ref{Fig2}a), confirming the AHE contribution for temperatures between 0.5 K and 10 K. This is illustrated in the inset of Fig.~\ref{Fig3}a for \textit{T} = 1.8 K, where the total Hall resistivity $\rho_{zx}$ (blue) is fit to Eq. (1) (red). Above $10~\text{K}$ (Supplementary Materials Fig. S5), the AHE contribution vanishes, resulting in $\rho_{zx}$ linear in field. The positive Hall resistivity slope indicates $p$-type behavior, and the carrier concentration $n_{\text{h}}$ and mobility $\mu_{\text{h}}$ are found to be $n_{\text{h}} = 7.8 \times 10^{22}\text{ cm}^{-3}$ and $\mu_{\text{h}} = 714.4\text{ cm}^2\text{ V}^{-1}\text{s}^{-1}$ at the lowest measured temperature. The Hall conductivity $\sigma_{zx}$ as a function of the magnetic field between \textit{T} = 0.5 K to 10 K is exhibited in Fig. \ref{Fig3}b. At lower temperatures $\sigma_{zx}$ increases rapidly and shows a linear increase with magnetic field. The linear increase shifts towards higher fields as the temperature increases. We obtained AHC ($\sigma^A_{zx}$) by extrapolating high-field Hall conductivity data to zero field as shown in the inset of Fig.~\ref{Fig3}b for \textit{T} = 1.8 K. The non-zero intercept persists up to 10 K, indicating that AHC is observed up to 10 K. 

To determine the nature of the anomalous Hall contribution (extrinsic or intrinsic), it is useful to estimate the AHC $\sigma^A_{zx}$ as a function of {$\sigma_{xx}$. $\sigma_{zx}$ as a function of the magnetic field between \textit{T} = 0.5 K to 10 K is shown in Fig. \ref{Fig3}b. The resulting anomalous contribution $\sigma^A_{zx}$ as a function of $\sigma_{xx}$ between 0.5 K to 10 K is shown in Fig. \ref{Fig3}c. At \textit{T} = 0.4 K, $\sigma^A_{zx}$ is found to be 160 S/cm and it remains nearly temperature-independent below $T_{\mathrm{K}} \approx 3.5~\text{K}.$ At higher temperatures, $\sigma^A_{zx}$ shows strong temperature dependence and reaches $\sim$ 20 S/cm around 10 K. The anomalous Hall angle $\theta^A$ is found to be $\theta^A~=~0.23^\circ$, similar to that of most magnetic compounds where $\theta^A~<~3^\circ$ \cite{yang2025modulation}. A comparison of $\sigma^{A}$ \textit{vs.} $\sigma_{xx}$ for CeAuAl$_3$ (blue) with other Ce-based compounds (black) (Fig. \ref{Fig3}d), places CeAuAl$_3$ within the intrinsic regime. Furthermore, the values of $\sigma^A_{zx}$ for CeAuAl$_3$ are comparable to those of the antiferromagnets Ce$_2$CuGe$_6$ and Ce$_2$PdGe$_6$ \cite{kotegawa2024large}, and even to those of some ferromagnets such as CeAlSi \cite{alam2023sign} and CePd$_2$P$_2$ \cite{li2025extrinsic}. The intrinsic anomalous Hall conductivity for CeAuAl$_3$ provides experimental confirmation of the topological features observed in the theoretical calculations.

To understand the microscopic origin of $\sigma^A_{zx}$ in CeAuAl$_3$, we performed detailed calculations of the electronic band structure using density functional theory (DFT). The three-dimensional (3D) Brillouin zone (BZ) is shown in the inset of Fig. \ref{Fig4}a. The PM band structure with SOC reveals KNLs along $\mathit{\Gamma}\text{-}Z$ and $X\text{-}P$, as shown in Fig.~\ref{Fig4}a. These bands are doubly-degenerate, and this degeneracy cannot be lifted by SOC, as expected for KNLs. These KNLs are protected by time-reversal symmetry along with achiral point group symmetries (mirror symmetries). Furthermore, in the presence of strong correlations, we find that the $\text{Ce}\ 4f$ states play a crucial role in pinning non-topological bands near the Fermi level $E_F$. Including these $\text{Ce}\ 4f$ states moves the KNLs close to the Fermi energy (as shown in the Supplementary Materials Fig.~S6a).

In contrast, band structure calculations in the spin-polarized state (Fig. \ref{Fig4}b) break time-reversal symmetry, lifting the degeneracy via a Zeeman field and giving rise to a field-induced AHE. The intrinsic anomalous Hall conductivity $\sigma^A_{int}$ originates from gapped KNLs, which generate substantial Berry curvature. Consistent with our model calculations (Fig. \ref{Fig1}b,c), DFT results confirm that these gapped KNLs drive the observed magnetotransport properties.

Using this spin-polarized framework, we calculated the AHC as a function of the position of the Fermi level to directly compare the DFT results (Fig.~\ref{Fig4}c) with our transport data. For an in-plane magnetization configuration ($\textit{H} \parallel {b}$), the calculated \(\sigma^A_{zx}\) at $E_F$ shows a large, substantial peak, while \(\sigma^A_{xy}\) and \(\sigma^A_{yz}\) remain negligible. Most importantly, the peak value of our calculated \(\sigma^A_{zx}\) at $E_F$ reaches approximately $220\text{~S/cm}$, which is in qualitative agreement with experimental value of \(\sigma^A_{zx}\) at the lowest temperature. Note that DFT calculations indicate a negligible anomalous Hall conductivity $\sigma_{xy}$. This is consistent with our magnetotransport measurements, which show no AHE contribution to the Hall resistivity $\rho_{yx}$ and exhibit a strictly linear field dependence for \textit{I} $||$ \textit{a} and \textit{H} $||$ \textit{c} (Supplementary Materials Fig.~S4c).

\begin{figure*}[!t]
\centering
\includegraphics[width=1\textwidth]{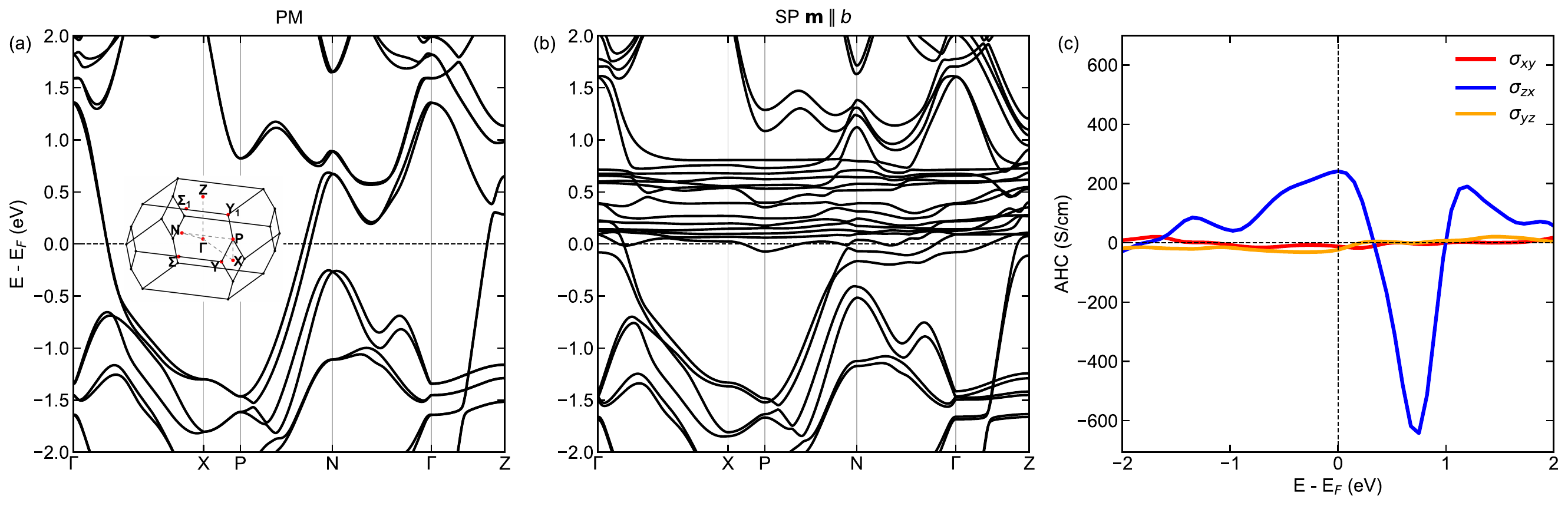}
\caption{\small \textbf{Electronic structure of CeAuAl$_3$.}
{(a) Density functional theory (DFT) calculation of the bulk electronic structure of CeAuAl$_3$ with spin-orbit coupling (SOC) in the paramagnetic state (PM). Inset shows the 3D Brillouin Zone. (b) DFT calculation of the spin polarized (SP) state with \textit{H} $||$ \textit{b} axis (c) AHC as a function of $E - E_{\mathrm{F}}$}  for the magnetization configuration with \textit{H} $||$ \textit{b} axis.} 
\label{Fig4}
\end{figure*}

\section{Discussion}

The AHE arises from an intrinsic contribution related to the momentum-space Berry curvature of the electronic bands and an extrinsic contribution due to the scattering of magnetic impurities \cite{nagaosa2010anomalous,berger1970side,xiao2010berry}. The origin of AHC can be understood by analyzing the scaling relationship between AHC and other physical quantities such as longitudinal conductivity $\sigma_{xx}$. The extrinsic skew scattering contribution shows the quadratic relationship between AHC and $\sigma_{xx}$ $(\sigma^{A}_{zx} \sim \sigma^{2}_{xx})$ \cite{tian2009proper}, while the intrinsic contribution due to the Berry curvature is independent of temperature and $\sigma_{xx}$ \cite{nagaosa2010anomalous,fang2003anomalous,haldane2004berry,xiao2010berry}.  The dominance of the extrinsic contribution has been widely discussed for various heavy-fermion compounds \cite{hadvzic1986hall,bogach2003hall,li2025extrinsic,kotegawa2024large,shen2026giant}, while the intrinsic contribution from the Berry curvature has been exhibited for a limited number of Kondo lattice compounds \cite{siddiquee2023breakdown,cheng2023observation,asaba2021colossal,xu2024tunable}. For CeAuAl$_3$, $\sigma^{A}_{zx}$ varies with $\sigma_{xx}$, reflecting the dominant extrinsic skew scattering for \textit{T} = 10 K to 3.5 K (Fig. \ref{Fig3}c). However, below the Kondo coherence temperature ($T_{\text{K}} = 3.5~\text{K}$), $\sigma^{A}_{zx}$ remains almost constant down to 0.5 K. This temperature-independent plateau (Fig. \ref{Fig3}c) signifies that intrinsic Berry curvature contributions dominate below $T_{\text{K}}$, producing an intrinsic $\sigma^A_{zx}~\approx$ 160 S/cm. This intrinsic value quantitatively matches the DFT calculations and can be attributed to Berry curvature engendered by the  KNLs.

DFT and model calculations indicate that the observed KNLs originate from the renormalized bands induced by the Kondo hybridization between \textit{f} and conduction electrons, giving rise to KNLs located in the vicinity of the Fermi energy, which, in turn, give rise to significant Berry curvature contribution. These observations indicate that the magnetotransport properties in CeAuAl$_3$ are engineered through a non-trivial band topology. Our resistivity measurements suggest the presence of Kondo lattice coherence in CeAuAl$_3$ below $T_K = 3.5 ~K$. Thermodynamic and transport measurements show that CeAuAl$_3$ is a moderate heavy-fermion compound. Therefore, in the presence of Kondo-lattice coherence, CeAuAl$_3$ is the first KNL Kondo metal. Nevertheless, a systematic exploration of ARPES is worthwhile for future investigations of the experimental band structure, although the best ARPES instruments still lack the ultrahigh energy resolution required to resolve details within the extremely narrow heavy-fermion bands. Furthermore, this system can be driven towards quantum criticality via chemical doping, pressure, or strain. Our findings demonstrate the interplay between KNLs topology, magnetic order, and time reversal symmetry.


\section{Methods}\label{S2}

\subsection{Crystal Growth and Structural Characterization}

Single crystals of CeAuAl$_3$ were grown by the flux method.  The starting elements with compositions Ce:Au:Al = 1:2:15 were placed in 2 ml alumina crucible and sealed in a fused silica ampule under vacuum. The sealed ampule was heated to 1100 $^{\circ}$C for 10 hours, homogenized at 1100 $^{\circ}$C for 4 hours, and then cooled to 1000 $^{\circ}$C at 1.66 $^{\circ}$C/h, when the excess flux was decanted using a centrifuge. Many well-formed rectangular single crystals were obtained, with typical sample dimensions of $2\text{ mm} \times 2\text{ mm} \times 0.5\text{ mm}$. The elemental composition was confirmed by energy dispersive x-ray spectroscopy (EDS). The crystal structure was determined using powder x-ray diffraction in a Bruker D8 Advance x-ray diffractometer with Cu K$_\alpha$ radiation. The Rietveld refinement shown in Supplementary Materials Fig. S1a was performed using the FULLPROF software.

\subsection{Resistivity, Specific heat and magnetotransport measurements}

Resistivity, \textit{MR}, Hall resistivity and heat capacity measurements were performed in a Quantum Design Dynacool Physical Property Measurement System (PPMS) with a 14 T magnet. The Magnetic Property Measurement System (MPMS) with a 7 T magnet was used for magnetization and magnetic susceptibility measurements. Single crystals of CeAuAl$_3$ were trimmed to adequate dimensions for electrical transport measurements. Crystals were oriented with the  [001], [100], and [120] directions parallel to the applied field for the \textit{c} axis, \textit{a} axis and \textit{b} axis  measurements. Resistivity and Hall measurements were performed using the 4-probe method. Pt wires of 12.5 $\mu\text{m}$ were used for electrical contacts with contact resistances $<$ 3 $\Omega$. Contacts were affixed with Epotek H20E silver epoxy. Magnetotransport and thermodynamic measurements were verified in more than one crystal for data reproducibility.

The samples were polished to dimensions of   $\sim 1.00\text{ mm} \times 0.40\text{ mm} \times 0.10\text{ mm}$, with the long dimension along \textit{a} axis. Contact misalignment in the \textit{MR} and Hall resistivity measurements was accounted for by using symmetrization and anti-symmetrization techniques using:

\begin{equation*}
  \rho_{ii}=\frac{\rho_{ii}(+H)+\rho_{ii}(-H)}{2}, \quad \rho_{ij}=\frac{\rho_{ij}(+H)-\rho_{ij}(-H)}{2}
\end{equation*}

Hall conductivity was calculated using the following relation:

\begin{equation*}
\sigma_{zx}
= -\frac{\rho_{zx}}{\rho_{zx}^2 + \rho_{xx}\rho_{zz}}
\end{equation*}

and, assuming $\rho_{xx}\sim\rho_{zz}$ gives

\begin{equation*}
\sigma_{zx}
= -\frac{\rho_{zx}}{\rho_{zx}^2 + \rho^2_{xx}}.
\end{equation*}

Here, both $\rho_{zx}$ and $\rho_{xx}$ were measured on the same sample.

\subsection{Density functional theory calculations}

DFT calculations were performed using the full-potential (linearized) augmented-plane-wave method, as implemented in WIEN2k~\cite{blaha2001wien2k}, and the projector augmented-wave (PAW) method as implemented in the Vienna ab \textit{initio} simulation package (VASP)~\cite{kresse1993ab,kresse1999ultrasoft}. In the paramagnetic phase, we treated the Ce $4f$ orbitals as core electrons. The generalized gradient approximation (GGA) of the exchange-correlation functional~\cite{perdew1996generalized} was used and spin-orbit coupling was included in the self-consistency cycles. A plane-wave cutoff of R$_{MT}$K$_{max}=9.5$ with muffin tin radii R$_{MT}$ set to 2.21 $a_0$ for Ce, 2.30 $a_0$ for Au and 2.00 $a_0$ for Al were employed. The anomalous Hall conductivity analysis was performed by employing a real-space tight-binding model Hamiltonian, which was obtained by using the wien2wannier~\cite{kunevs2010wien2wannier,pizzi2020wannier90} interface and analyzed using wanniertools~\cite{wu2018wanniertools}. The helical magnetic structure was calculated using VASP where the Ce $4f$ state were treated as valence states.

\section{Acknowledgments}
We would like to thank Pavan Hosur for valuable discussions. The experimental work at Rice University has been primarily supported by the Vannevar Bush Faculty Fellowship ONR-VB N00014-24-1-2048 (HB, A, SM, EM) and the Robert A. Welch Foundation Grant No. C-2114 (KR, EM). We also acknowledge partial support from the US DOE BES DE-SC0026179 (EM, KR, QS). The theory work from Rice has
been primarily supported by the NSF Grant No. DMR2220603 (YF, KSL), the Robert A. Welch Foundation
Grant No. C-1411 (MM) and the Vannevar Bush Faculty
Fellowship ONR-VB N00014-23-1-2870 (MM, QS). K.-S.L. acknowledges the Carl and
Lillian Illig Postdoctoral Fellowship from the SmalleyCurl Institute at Rice University. The computational
calculations have in part been performed on the Shared
University Grid at Rice funded by NSF under Grant
EIA-0216467, a partnership between Rice University, Sun
Microsystems, and Sigma Solutions, Inc., the Big-Data
Private-Cloud Research Cyberinfrastructure MRI-award
funded by NSF under Grant No. CNS-1338099, and
the Extreme Science and Engineering Discovery Environment (XSEDE) by NSF under Grant No. DMR170109.
This research used resources of the National Energy Research Scientific Computing Center (NERSC), a DOE
Office of Science User Facility supported by the Office of Science of the U.S. Department of Energy under contract no. DE-AC02-05CH11231 using NERSC
award BES-ERCAP0020966. The work at Los Alamos
National Laboratory (LANL) was carried out under the
auspices of the U.S. Department of Energy (DOE) National Nuclear Security Administration under Contract
No. 89233218CNA000001, and was supported by the
LANL Laboratory Directed Research and Development
Program, and in part by the Center for Integrated Nanotechnologies, an Office of Science User Facility operated
by the DOE Office of Science, in partnership with the
LANL Institutional Computing Program for computational resources.

\section{Author Contribution}
H.B. and E.M. conceived the idea and coordinated the project. H.B. synthesized the single crystals, performed magnetic and magnetotransport measurements, and analyzed the data. K.A. performed the DFT band structure and AHC calculations. Y.F., K.-S.L., M.M., and Q.S. performed the symmetry analysis and toy-model calculations and provided input for the theoretical discussion. A., K.R., and S.M. assisted with magnetotransport measurements and data analysis. C.W. and J.-X.Z. provided input for the DFT band structure calculations. H.B. and E.M. wrote the manuscript with input from all the authors.

\section{Competing interests}
The authors declare no Competing Financial or Non-Financial Interests.

\section{Data availability}
The authors declare that the main data supporting the findings of this study are available in the article and the supplementary material. Additional data is available from the corresponding author upon request.

\clearpage
\newpage

\section*{References}

\widetext
\begin{center}
\pagebreak
\hspace{0pt}
\vfill
\textbf{\large Supplementary Material: Discovery of Kondo-Kramers Nodal Line Topological State in strongly correlated material CeAuAl$_3$}
\vfill
\hspace{0pt}
\end{center}

\setcounter{equation}{0}
\setcounter{figure}{0}
\setcounter{table}{0}
\setcounter{page}{1}
\makeatletter
\renewcommand\thesection{SM\arabic{section}}
\renewcommand{\theequation}{S\arabic{equation}}
\renewcommand{\thetable}{S\arabic{table}}
\renewcommand\thefigure{S\arabic{figure}}
\renewcommand{\theHtable}{S\thetable}
\renewcommand{\theHfigure}{S\thefigure}

\maketitle

\section*{S1.~~Structural characterization}

\begin{figure*}[!ht]
\begin{center}
\includegraphics[width=1\linewidth]{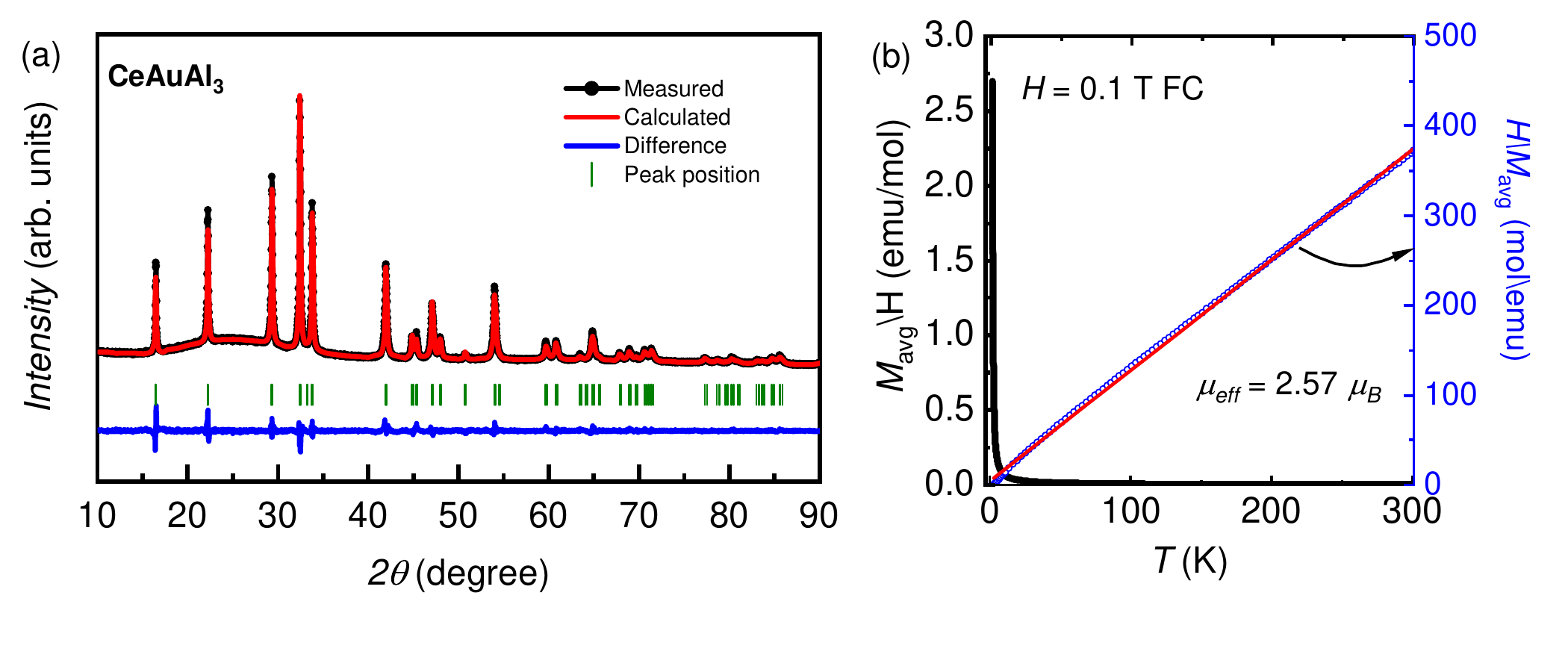}
    \caption{\small (a) Rietveld refinement of room temperature powder x-ray diffraction pattern of CeAuAl$_3$ (b) Average susceptibility $M_{\text{avg}}/H$ (left panel), and inverse susceptibility, $H/M_{\text{avg}}$ (right panel), as a function of temperature from $1.8\text{--}300\text{ K}$. }
    \label{S1}
    \end{center}
\end{figure*}

The powder X-ray diffraction pattern along with its Rietveld refinement are shown in Fig.~\ref{S1}a (black symbols and red line, respectively). Figure~\ref{S1}b illustrates the averaged magnetic susceptibility $M_{avg}/H$ along with its inverse $H/M_{avg}$, where $M_{avg} = \frac{1}{3}(2M_a + M_c)$. The high temperature magnetization follows the Curie-Weiss law with an effective magnetic moment $\mu_{\text{eff}} \approx 2.57 \, ~\mu_B$ and a Curie-Weiss temperature  $\theta_p \approx -4.63\,\text{K}$. The obtained $\mu_{\text{eff}}$ is in good agreement with the theoretical value for a free $Ce^{3+}$ ion ($2.54\,~ \mu_B$). The negative value of $\theta_p$ suggests that AFM interactions are dominant in the system.

\section*{S2.~~Resistivity}

 \begin{figure*}[!ht]
\begin{center}
\includegraphics[width=1\linewidth]{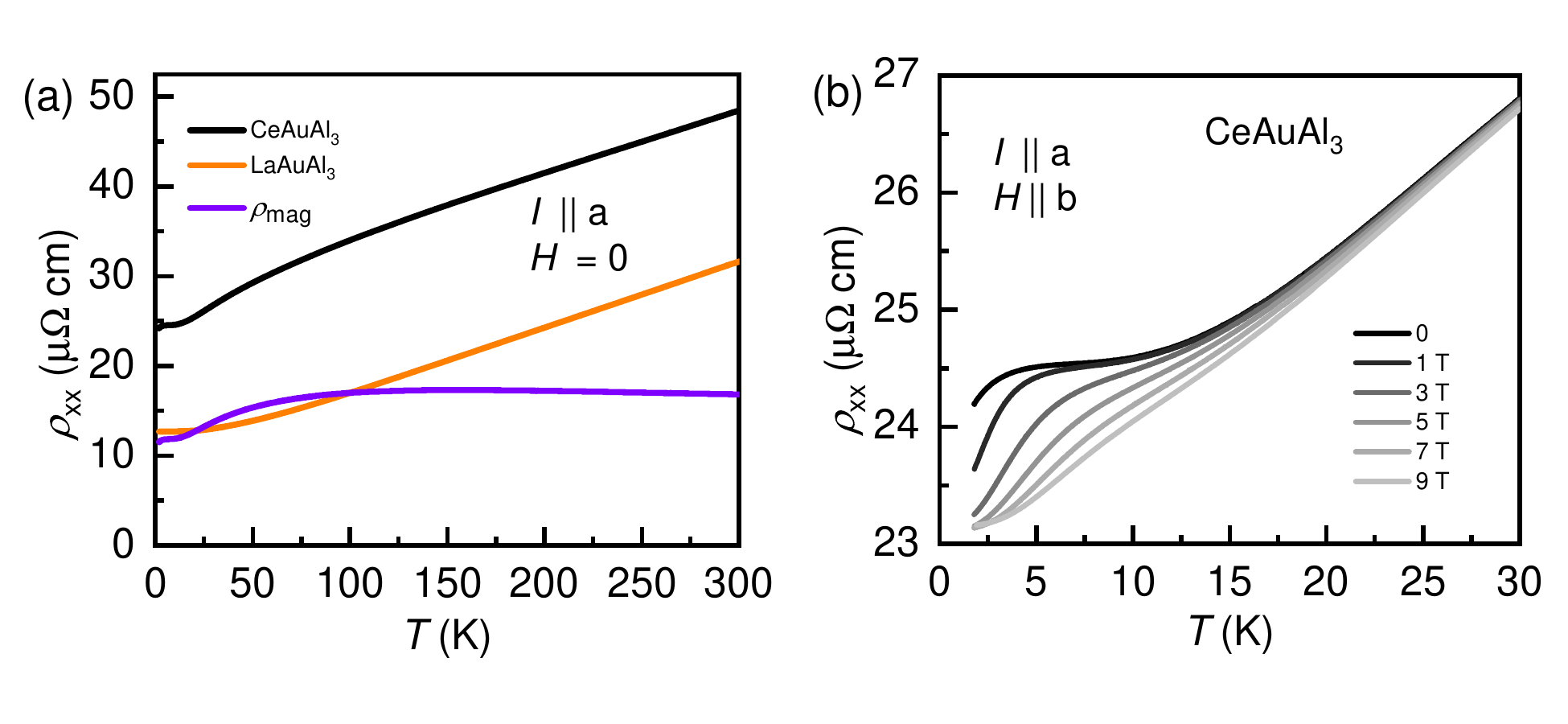}
    \caption{\small (a). Temperature dependent resistivity of CeAuAl$_3$ with \textit{I} $||$ \textit{a}. The resistivity of LaAuAl$_3$ is shown by orange curve and the magnetic contribution is shown by purple curve. (b) Temperature dependent electrical resistivity as a function of temperature with current \textit{I} $||$ \textit{a} and \textit{H} $||$ \textit{b} with different fields. }
    \label{S2}
    \end{center}
\end{figure*}

The resistivity as a function of temperature for CeAuAl$_3$ and LaAuAl$_3$ is shown in Fig.~\ref{S2}a. $\rho_{xx}$ of LaAuAl$_3$ displays a typical metallic profile, whereas that of CeAuAl$_3$ exhibits a slight enhancement at low temperature consistent with Kondo-like behavior. $\rho_{xx}$ for CeAuAl$_3$ shows a slight upturn around \textit{T} = 10 K and a shoulder around \textit{T} = 3.5 K indicating the incoherent single-ion Kondo scattering temperature and the Kondo coherence temperature respectively, while these features are absent in LaAuAl$_3$ data.  The rapid decrease of resistivity below the maximum indicates the onset of long range magnetic order. The magnetic contribution to the resistivity indicated by the purple curve shows a hump around \textit{T} = 50 K as a result of the scattering from an excited crystal electric field (CEF) level. The upturn and $\rho_{\text{max}}$ are not particularly prominent---indicating a weak Kondo effect---though they become visible after subtracting the resistivity of La. Fig.~\ref{S2}b indicates the resistivity of CeAuAl$_3$ measured with magnetic fields up to \textit{H} = 9 T. The application of external fields causes an overall suppression of the low temperature electrical resistivity, with the coherence peak shifting to higher temperatures as the field increases \cite{kawakami1986magnetoresistance,parihari2009magnetoresistance}. These observations indicate typical features of Kondo lattices in which the Kondo interaction is quenched by the external magnetic field.

\newpage

\section*{S3.~~Heat Capacity}

 \begin{figure*}[!ht]
\begin{center}
\includegraphics[width=1\linewidth]{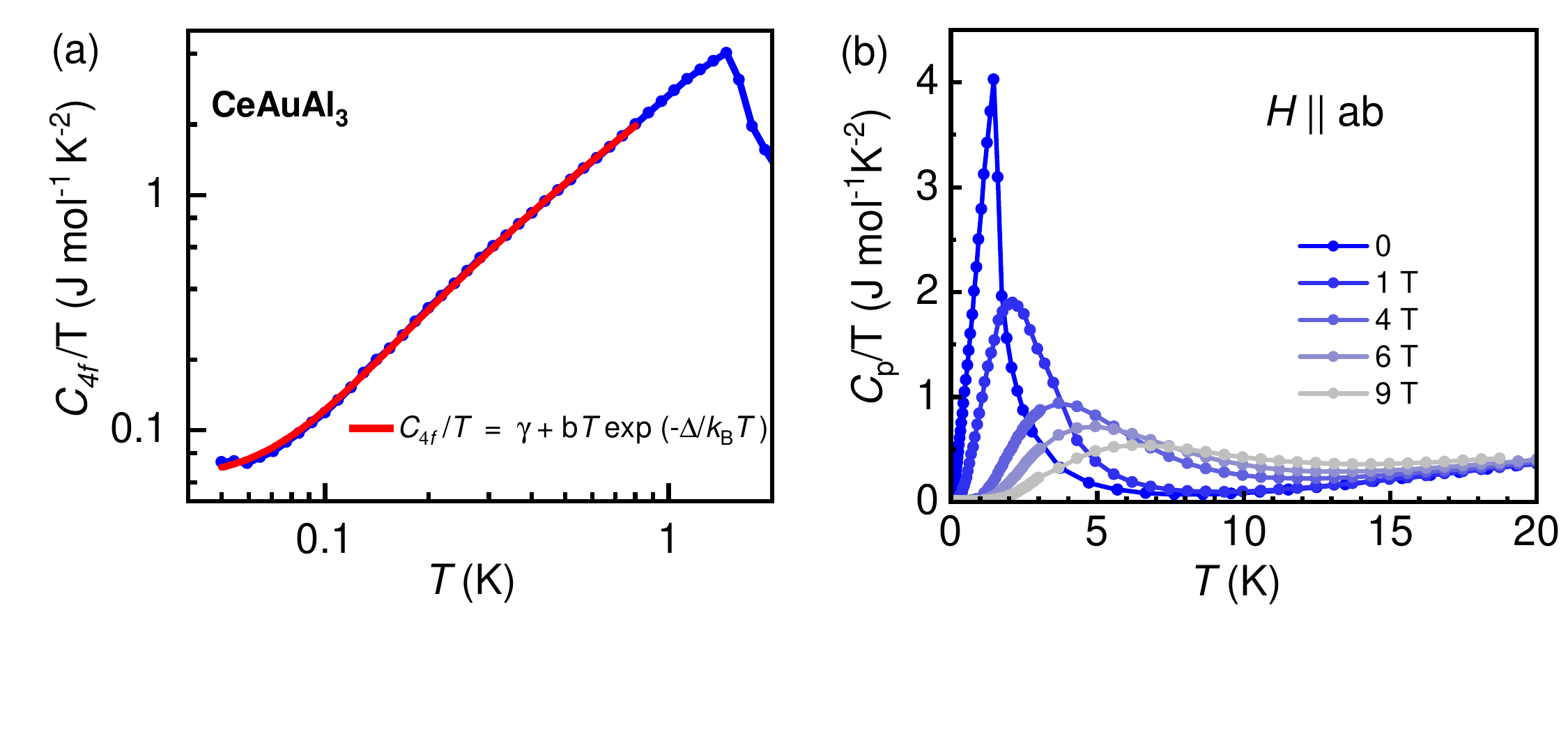}
    \caption{\small (a) $C_{4f}/T$ as a function of temperature plotted below 2 K. Red curve indicates fit to $C_{4f}/T$ data (b). Heat Capacity $C_p$/T as a function of temperature with different fields from $H$ = 0 to 9 T. }
    \label{S3}
    \end{center}
\end{figure*}

At zero field, the specific heat $C_{p}/T$ exhibits a sharp $\lambda$-type anomaly at $T_\mathrm{N}$ $\approx$ 1.3\text{ K}, signaling a long-range magnetic transition into an incommensurate helical state as shown in Fig.~\ref{S3}a. Blue curve isolates the 4f-electronic contribution \textit{C}$_{4f}$ to the specific heat of CeAuAl$_3$ in zero field and fit of the data to $\frac{C_{4f}}{T} = \gamma + b \, T \, \exp\left(-\frac{\Delta}{k_B T}\right)$ indicated by red curve. The first term represents Sommerfield coefficient of the heat capacity and is found to be 73 mJ/mole~K$^{2}$. The second term  $b \, T \, e^{-\Delta/(k_B T)}$ represents AFM magnon term which arises from an AFM spin-density with a gap in the excitation spectrum due to anisotropy \cite{cornelius2000anisotropic}, and suggests the presence of gapped spin waves with $\frac{\Delta}{k_B}$ = 160 mK, where $\Delta$ represents the activation energy. This equation provides a good fit of the CeAuAl$_3$ data below 0.5 $T_{N}$ as shown by the red curve in Fig.~\ref{S3}a. Specific heat provides signatures of heavy fermion behavior and magnetic order, indicating the mass enhancement of quasiparticles even in the magnetically ordered state. The $\lambda$-type anomaly observed at zero field broadens with the magnetic field applied in the $ab$-plane, as shown in Fig.~\ref{S3}b.  The Schottky anomaly emerges and systematically shifts to higher temperatures as the field increases. This field induced Schottky behavior arises due to Zeeman splitting of the doublet CEF ground state.

\section*{S4.~~Magnetotransport}

 \begin{figure*}[ht]
\begin{center}
\includegraphics[width=1\linewidth]{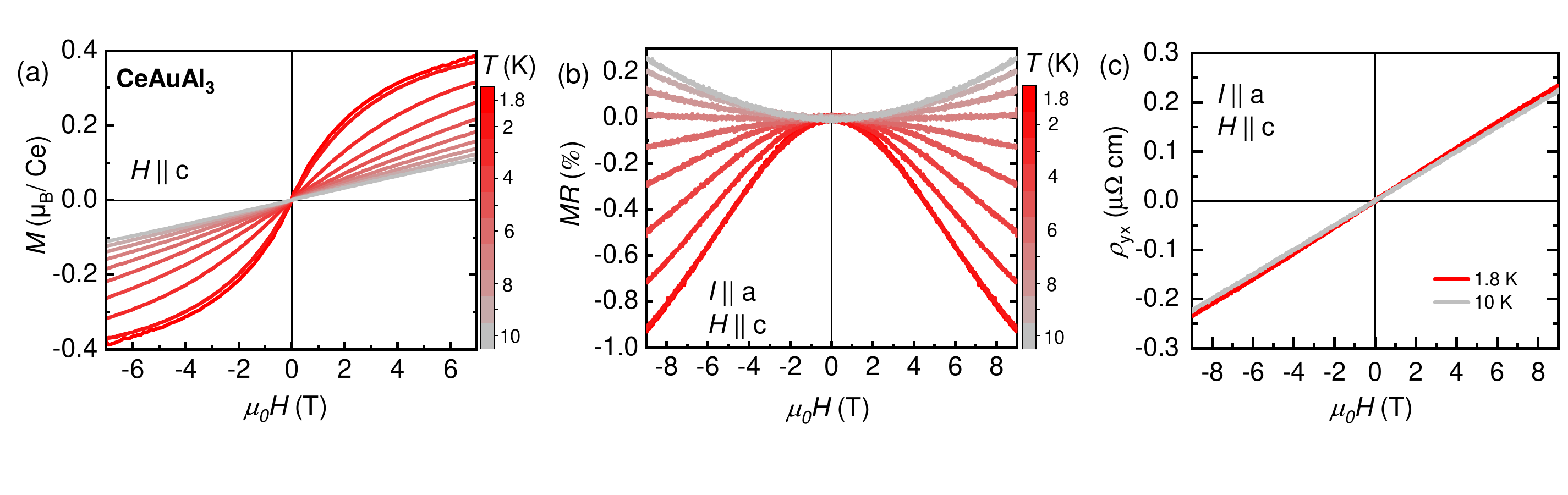}
    \caption{\small (a) Magnetization as a function of magnetic field with $H \parallel c$ measured from 1.8--10~K. (b).  Magnetoresistance as a function of magnetic field for \textit{I} $||$ \textit{a} and \textit{H} $||$ \textit{c} measured between 1.8-10 K. (c) Hall resistivity ($\rho_{yx}$) as a function of magnetic field for \textit{H} $||$ \textit{c} and \textit{I} $||$ \textit{a} for \textit{T} = 1.8 K and 10 K.}
    \label{S4}
    \end{center}
\end{figure*}

Supplementary Fig.~\ref{S4}a shows magnetization as a function of magnetic field between \textit{T} = 1.8 K to 10 K for \textit{H} $||$ \textit{c} axis. At \textit{T} = 1.8 K, the curve $M(H)$ increases linearly with the field and rises without showing any sign of saturation. This behavior is persistent at lower temperatures, and at higher temperatures $M(H)$ curve becomes linear. The most noticeable consequences of Kondo correlations is observed in \textit{MR} measured as a function of the magnetic field for current \textit{I} $||$ a and magnetic field \textit{H} $||$ c between 1.8 K and 10 K, as shown in Fig.~\ref{S4}b. At temperatures above the resistivity minimum ($T > 8~\mathrm{K}$, main text Fig.~1f), the $MR$ (gray curves) is positive, as expected for a normal metal. The negative \textit{MR} at lower temperatures is most likely from coherent scattering \cite{kawakami1986magnetoresistance}. The onset of incoherent scattering below \textit{T} = 8 K is indicated by negative/positive  \textit{MR} observed below/above 8 K, and is consistent with the behavior observed for other Ce-based heavy fermion compounds \cite{rauchschwalbe1987magnetoresistance}. The observed \textit{MR} for field applied along the hard axis clearly distinguishes features observed as a result of incoherent and coherent scattering. Supplementary Fig.~\ref{S4}c shows the Hall resistivity ($\rho_{yx}$) as a function of the magnetic field for \textit{H} $||$ \textit{c} and \textit{I} $||$ \textit{a} for \textit{T} = 1.8 K and 10 K. $\rho_{yx}$ exhibit a linear behavior indicating the ordinary Hall effect. 

 \begin{figure*}[ht]
\begin{center}
\includegraphics[width=1\linewidth]{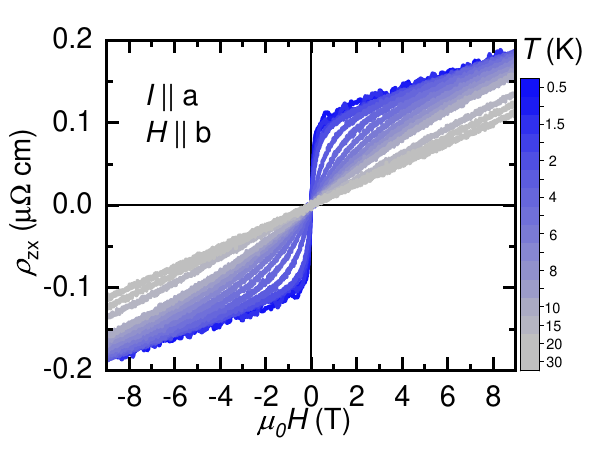}
    \caption{\small (a) Hall resistivity ($\rho_{zx}$) as a function of magnetic field for \textit{H} $||$ \textit{b} and \textit{I} $||$ \textit{a} for selected temperatures between \textit{T} = 0.5-30 K.}
    \label{S5}
    \end{center}
\end{figure*}

\newpage

\section*{S5.~~Density functional theory calculations}

 \begin{figure*}[!ht]
\begin{center}
\includegraphics[width=1\linewidth]{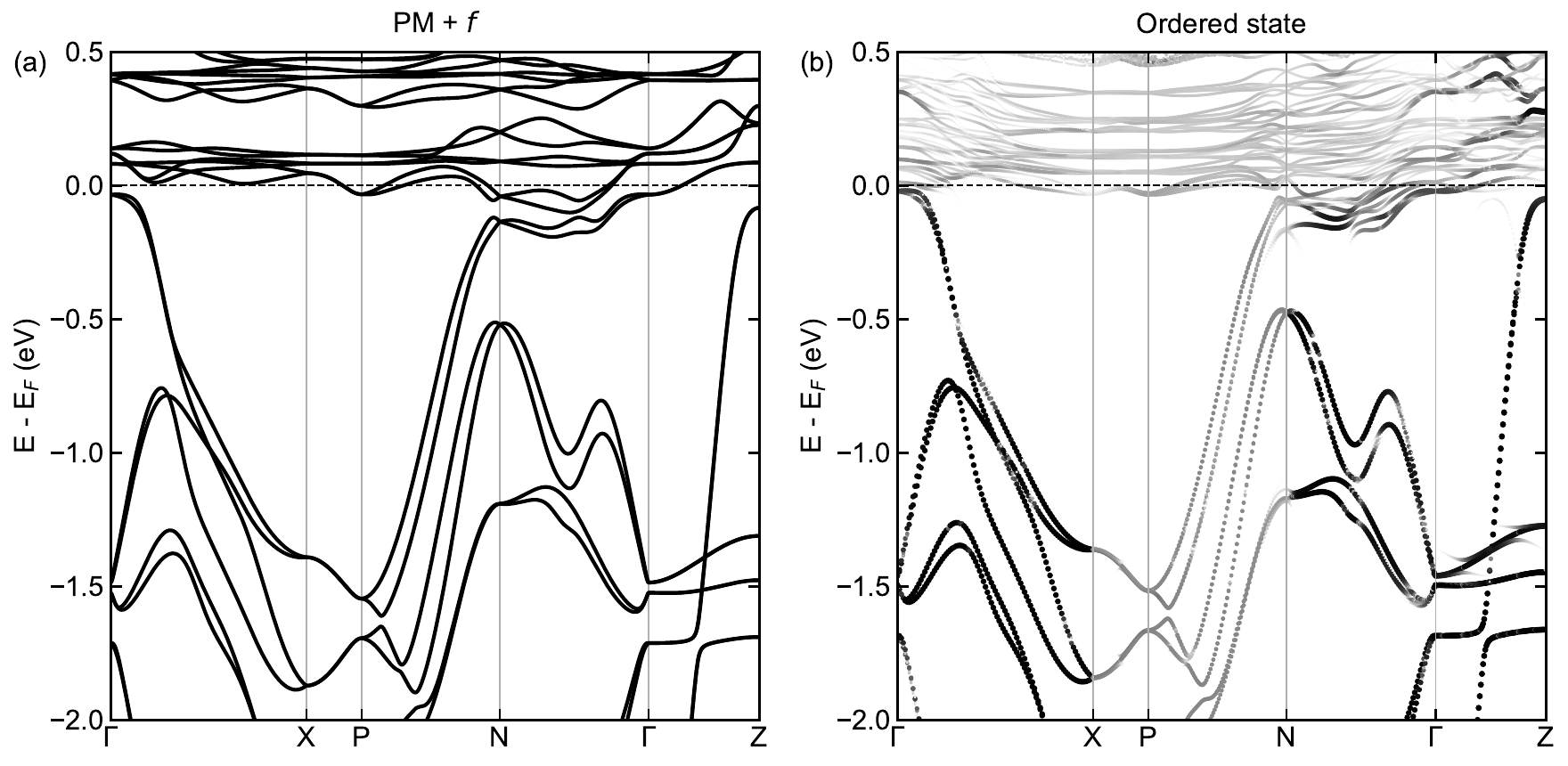}
    \caption{\small (a). DFT calculations for electronic structure of CeAuAl$_3$ with SOC including 4f in PM state (b) ordered state.}
    \label{S6}
    \end{center}
\end{figure*}

\newpage

\section*{S6.~~Toy Model and Parton Calculation}

We construct a toy model with two orbitals at the $1a$ Wyckoff position, with coordinates $\mathbf{r} = (0,0,z)$. Each orbital has two spin degrees of freedom. The two orbitals correspond to two elementary band representations (EBRs), which we denote by $c_{1\sigma}$ and $c_{2\sigma}$. We organize them into the basis $\Psi^\dagger = (\Psi^\dagger_1, \Psi^\dagger_2)$ and $\Psi^\dagger_I = (c^\dagger_{I\uparrow}, c^\dagger_{I\downarrow})$, where $I = 1,2$ labels the EBR. In this basis, the symmetry-allowed independent hoppings can be systematically determined~\cite{po2017symmetry, bradlyn2019disconnected, lin2026chiral}.

The noninteracting Hamiltonian for the conduction electrons that preserves crystalline and time-reversal symmetries takes the form
\begin{equation}
H_0 = \sum_{\mathbf{k}} \Psi^\dagger_{\mathbf{k}} h_0(\mathbf{k}) \Psi_{\mathbf{k}}, \quad h_0(\mathbf{k}) = 
\begin{pmatrix} 
h_{11} & h_{12} \\ 
h^\dagger_{12} & h_{22} 
\end{pmatrix}.
\end{equation}

We consider nearest-neighbor and next-nearest neighbor hopping terms and spin-orbit coupling (SOC). Writing $k_x$ and $k_y$ in units of $1/a$ and $k_z$ in units of $1/c$, then each block matrix is given by:

\begin{align}
h_{II}(\mathbf{k}) = & \left[p^I_0 - 8p^I_1 \cos\frac{k_x}{2} \cos\frac{k_y}{2} \cos\frac{k_z}{2}\right] \sigma_0 \nonumber \\
& - 4p^I_2 \sin\frac{k_y}{2} \cos\frac{k_x}{2} \cos\frac{k_z}{2} \sigma_x + 4p^I_2 \sin\frac{k_x}{2} \cos\frac{k_y}{2} \cos\frac{k_z}{2} \sigma_y \nonumber \\
& + 2p^I_4 [\cos k_x + \cos k_y] \sigma_0 + 2p^I_5 [\sin k_y \sigma_x - \sin k_x \sigma_y]
\end{align}

\begin{equation}
h_{12}(\mathbf{k}) = m \sigma_0.
\end{equation}

Here $p^I_a$ denotes a hopping parameter and $m$ is an onsite hybridization strength. Specifically, $p^I_0$ is the onsite energy; $p^I_1$ is the nearest-neighbor intra-orbital hopping amplitude; $p^I_2$ is the nearest-neighbor intra-orbital SOC amplitude. Similarly, $p^I_4$ is the next-nearest-neighbor intra-orbital hopping amplitude; $p^I_5$ is the next-nearest-neighbor intra-orbital SOC amplitude. These SOC terms are the lattice version of the Rashba SOC. The parameter values used in our calculations are summarized in Table~\ref{tab:parameters}.

\begin{table}[b]
\caption{\label{tab:parameters} Parameter values of the toy model and the corresponding saddle-point solutions. All quantities are in units of the nearest-neighbor hopping amplitude $t$ of the second conduction electron orbital, defined by $|p^2_1| = t$.}
\begin{ruledtabular}
\begin{tabular}{ccccccc}
$I$ & $p^I_0$ & $p^I_1$ & $p^I_2$ & $p^I_4$ & $p^I_5$ & $m$ \\ \hline
1 & $-6.0$ & $-0.2$ & $0.5$ & $1.0$ & $0.2$ & $1.0$ \\
2 & $7.0$  & $-1.0$ & $0.5$ & $1.0$ & $0.2$ & \\ 
\hline
$\epsilon_f$ & $V_1$ & $V_2$ & $\nu$ & $r$ & $\lambda$ & $\mu$ \\ \hline
$-1.00$ & $1.0$ & $2.0$ & 3 & $0.205$ & $1.755$ & $0.729$ \\
\end{tabular}
\end{ruledtabular}
\end{table}

We further introduce a localized $f$ orbital with Hubbard interaction $U$, thereby forming a periodic Anderson lattice model. The $f$ orbitals couple to the two conduction electrons through Kondo hybridization with strength $V_I$. The Hamiltonian is

\begin{equation}
H = H_0 + \sum_{i,\sigma} \left[ \epsilon_f n_{f,i\sigma} + \sum_{I=1,2} V_I \left(f^\dagger_{i\sigma} c_{I,i\sigma} + \text{h.c.}\right) \right] + \sum_{i} U n_{f,i\uparrow} n_{f,i\downarrow} - \mu \sum_{i,\sigma} (n_{c_1,i\sigma} + n_{c_2,i\sigma} + n_{f,i\sigma})
\end{equation}

where $i$ is the site label representing the unit cell label, and $f^\dagger_{i\sigma}$ is the creation operator for an $f$ electron at site $i$ with spin $\sigma$. Here $n_{f,i\sigma} = f^\dagger_{i\sigma} f_{i\sigma}$ is the $f$-electron number operator, and $\epsilon_f$ is the on-site energy of the $f$-electron. $\mu$ is the chemical potential and $n_{cI,i\sigma} = c^\dagger_{I,i\sigma} c_{I,i\sigma}$ is the conduction electron number operator for orbital $I$ at site $i$ with spin $\sigma$. In the infinite-$U$ limit, the number of $f$-electrons on each site is constrained by $\sum_{\sigma} \langle n_{f,i\sigma} \rangle \leq 1$. Hereafter, $\langle \cdots \rangle$ is understood as the ground state expectation value at zero temperature or the thermal average at finite temperature.

To treat the constraint, an auxiliary boson operator \textit{b} and a Lagrange multiplier $\lambda$ are introduced such that the periodic Anderson model in the infinite-$U$ limit in the auxiliary boson representation is

\begin{align}
H = H_0 & + \sum_{i,\sigma} \left[ \epsilon_f n_{f,i\sigma} + \sum_{I=1,2} V_I \left( (b_i f^\dagger_{i\sigma}) c_{I,i\sigma} + \text{$h.c.$} \right) \right] \nonumber \\
& + \lambda \sum_{i} \left( \sum_{\sigma} n_{f,i\sigma} + b^\dagger_i b_i - 1 \right) - \mu \sum_{i,\sigma} (n_{c_1,i\sigma} + n_{c_2,i\sigma} + n_{f,i\sigma})
\end{align}

where $\lambda$ is the Lagrange multiplier. At the saddle-point level, a solution can be obtained by considering the condensation of the auxiliary boson operator, which means effectively replacing $b_i$ by $\langle b \rangle$. We may absorb the complex phase of $\langle b \rangle$ by the $f$ electron operators such that we further replace $\langle b \rangle$ by $r$, where $r$ is a real number. This leads to the following self-consistent equations:

\begin{equation}
r = \sqrt{1 - \sum_{\sigma} \langle f^\dagger_{i\sigma} f_{i\sigma} \rangle},
\end{equation}
\begin{equation}
\lambda = \sum_{I=1,2} \frac{-V_I}{r} \sum_{\sigma} \text{Re} \langle c^\dagger_{I,i\sigma} f_{i\sigma} \rangle .
\end{equation}

Here we have assumed that the expectation values $\langle f^\dagger_{i\sigma} f_{i\sigma} \rangle$ and $\langle c^\dagger_{I,i\sigma} f_{i\sigma} \rangle$ for each site $i$ are identical due to symmetries. In addition, the chemical potential is determined by fixing the total particle number throughout the calculations:
\begin{equation}
\sum_{i,\sigma} \langle n_{c1,i\sigma} + n_{c2,i\sigma} + n_{f,i\sigma} \rangle = \nu N_{\text{L}},
\end{equation}
where $N_{\text{L}}$ is the total number of sites and $\nu$ is the electron filling. We set $\nu = 3$, which constrains each site to be half-filled. These equations can be solved self-consistently to get the saddle-point solution. The bare parameters $\epsilon_f, V_1, V_2$ and the self-consistent parameters $r, \lambda, \mu$ are listed in Table~\ref{tab:parameters}. The saddle-point solution is shown in the main text Fig. 1c. 


To illustrate how an external magnetic field opens gaps for Kramers nodal lines, we add a small Zeeman term to all orbitals:
\begin{equation}
H_{\text{Zeeman}} = \sum_{i}\sum_{I=1,2} m_1 M\cdot \sigma_{ss'}  c^\dagger_{iIs} c_{iIs'} + \sum_{i} m_2 M\cdot \sigma_{ss'}f^\dagger_{is} f_{is'} 
\end{equation}
where $m_1$, $m_2$ are the effective Zeeman coupling strengths, $M=M(B,T)$ is the ordered moment, and we choose $m_1=m_2=m$ in this calculation. Full band structure of the saddle point solution is shown in Fig. \ref{S7}. Zoomed-in plot for saddle-point solutions with $mM=0$ and $mM = 0.05~T_{\mathrm{K}}$ is shown in Fig.~1 in the main text, where the small Zeeman term opens sizable gaps in the heavy-fermion bands. This sensitivity to a small Zeeman field is a consequence of the strongly renormalized energy scale of the heavy-fermion bands.

\begin{figure*}[!ht]
\begin{center}
\includegraphics[width=0.8\linewidth]{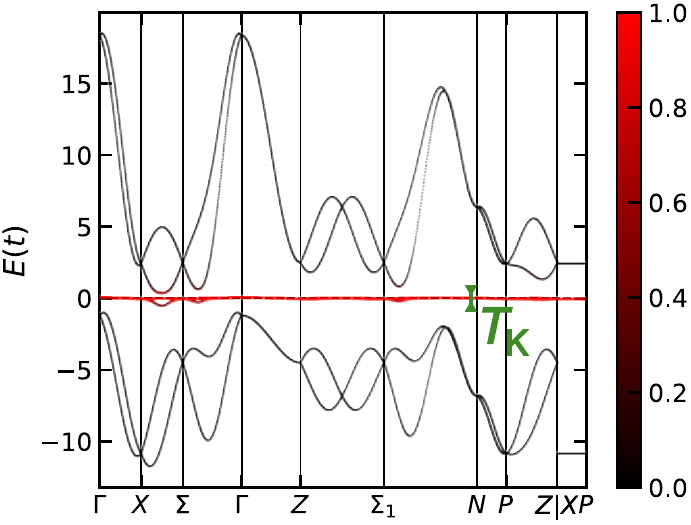}
    \caption{\small Full band structure of the saddle point solution for the toy model.}
    \label{S7}
    \end{center}
\end{figure*}


\begin{thebibliography}{67}%
\makeatletter
\providecommand \@ifxundefined [1]{%
 \@ifx{#1\undefined}
}%
\providecommand \@ifnum [1]{%
 \ifnum #1\expandafter \@firstoftwo
 \else \expandafter \@secondoftwo
 \fi
}%
\providecommand \@ifx [1]{%
 \ifx #1\expandafter \@firstoftwo
 \else \expandafter \@secondoftwo
 \fi
}%
\providecommand \natexlab [1]{#1}%
\providecommand \enquote  [1]{``#1''}%
\providecommand \bibnamefont  [1]{#1}%
\providecommand \bibfnamefont [1]{#1}%
\providecommand \citenamefont [1]{#1}%
\providecommand \href@noop [0]{\@secondoftwo}%
\providecommand \href [0]{\begingroup \@sanitize@url \@href}%
\providecommand \@href[1]{\@@startlink{#1}\@@href}%
\providecommand \@@href[1]{\endgroup#1\@@endlink}%
\providecommand \@sanitize@url [0]{\catcode `\\12\catcode `\$12\catcode `\&12\catcode `\#12\catcode `\^12\catcode `\_12\catcode `\%12\relax}%
\providecommand \@@startlink[1]{}%
\providecommand \@@endlink[0]{}%
\providecommand \url  [0]{\begingroup\@sanitize@url \@url }%
\providecommand \@url [1]{\endgroup\@href {#1}{\urlprefix }}%
\providecommand \urlprefix  [0]{URL }%
\providecommand \Eprint [0]{\href }%
\providecommand \doibase [0]{https://doi.org/}%
\providecommand \selectlanguage [0]{\@gobble}%
\providecommand \bibinfo  [0]{\@secondoftwo}%
\providecommand \bibfield  [0]{\@secondoftwo}%
\providecommand \translation [1]{[#1]}%
\providecommand \BibitemOpen [0]{}%
\providecommand \bibitemStop [0]{}%
\providecommand \bibitemNoStop [0]{.\EOS\space}%
\providecommand \EOS [0]{\spacefactor3000\relax}%
\providecommand \BibitemShut  [1]{\csname bibitem#1\endcsname}%
\let\auto@bib@innerbib\@empty
\bibitem [{\citenamefont {Paschen}\ and\ \citenamefont {Si}(2021)}]{paschen2021quantum}%
  \BibitemOpen
  \bibfield  {author} {\bibinfo {author} {\bibfnamefont {S.}~\bibnamefont {Paschen}}\ and\ \bibinfo {author} {\bibfnamefont {Q.}~\bibnamefont {Si}},\ }\bibfield  {title} {\bibinfo {title} {Quantum phases driven by strong correlations},\ }\href {https://www.nature.com/articles/s42254-020-00262-6} {\bibfield  {journal} {\bibinfo  {journal} {Nat. Rev. Phys.}\ }\textbf {\bibinfo {volume} {3}},\ \bibinfo {pages} {9} (\bibinfo {year} {2021})}\BibitemShut {NoStop}%
\bibitem [{\citenamefont {Checkelsky}\ \emph {et~al.}(2024)\citenamefont {Checkelsky}, \citenamefont {Bernevig}, \citenamefont {Coleman}, \citenamefont {Si},\ and\ \citenamefont {Paschen}}]{checkelsky2024flat}%
  \BibitemOpen
  \bibfield  {author} {\bibinfo {author} {\bibfnamefont {J.~G.}\ \bibnamefont {Checkelsky}}, \bibinfo {author} {\bibfnamefont {B.~A.}\ \bibnamefont {Bernevig}}, \bibinfo {author} {\bibfnamefont {P.}~\bibnamefont {Coleman}}, \bibinfo {author} {\bibfnamefont {Q.}~\bibnamefont {Si}},\ and\ \bibinfo {author} {\bibfnamefont {S.}~\bibnamefont {Paschen}},\ }\bibfield  {title} {\bibinfo {title} {Flat bands, strange metals and the {K}ondo effect},\ }\href {https://www.nature.com/articles/s41578-023-00644-z} {\bibfield  {journal} {\bibinfo  {journal} {Nat. Rev. Mater.}\ }\textbf {\bibinfo {volume} {9}},\ \bibinfo {pages} {509} (\bibinfo {year} {2024})}\BibitemShut {NoStop}%
\bibitem [{\citenamefont {Gegenwart}\ \emph {et~al.}(2008)\citenamefont {Gegenwart}, \citenamefont {Si},\ and\ \citenamefont {Steglich}}]{gegenwart2008quantum}%
  \BibitemOpen
  \bibfield  {author} {\bibinfo {author} {\bibfnamefont {P.}~\bibnamefont {Gegenwart}}, \bibinfo {author} {\bibfnamefont {Q.}~\bibnamefont {Si}},\ and\ \bibinfo {author} {\bibfnamefont {F.}~\bibnamefont {Steglich}},\ }\bibfield  {title} {\bibinfo {title} {Quantum criticality in heavy-fermion metals},\ }\href {https://www.nature.com/articles/nphys892} {\bibfield  {journal} {\bibinfo  {journal} {Nat. Phys.}\ }\textbf {\bibinfo {volume} {4}},\ \bibinfo {pages} {186} (\bibinfo {year} {2008})}\BibitemShut {NoStop}%
\bibitem [{\citenamefont {Coleman}\ and\ \citenamefont {Schofield}(2005)}]{coleman2005quantum}%
  \BibitemOpen
  \bibfield  {author} {\bibinfo {author} {\bibfnamefont {P.}~\bibnamefont {Coleman}}\ and\ \bibinfo {author} {\bibfnamefont {A.~J.}\ \bibnamefont {Schofield}},\ }\bibfield  {title} {\bibinfo {title} {Quantum criticality},\ }\href {https://www.nature.com/articles/nature03279} {\bibfield  {journal} {\bibinfo  {journal} {Nature}\ }\textbf {\bibinfo {volume} {433}},\ \bibinfo {pages} {226} (\bibinfo {year} {2005})}\BibitemShut {NoStop}%
\bibitem [{\citenamefont {Siddiquee}\ \emph {et~al.}(2023)\citenamefont {Siddiquee} \emph {et~al.}}]{siddiquee2023breakdown}%
  \BibitemOpen
  \bibfield  {author} {\bibinfo {author} {\bibfnamefont {H.}~\bibnamefont {Siddiquee}} \emph {et~al.},\ }\bibfield  {title} {\bibinfo {title} {Breakdown of the scaling relation of anomalous {H}all effect in {K}ondo lattice ferromagnet {USbTe}},\ }\href {https://www.nature.com/articles/s41467-023-36221-9} {\bibfield  {journal} {\bibinfo  {journal} {Nat. Commun.}\ }\textbf {\bibinfo {volume} {14}},\ \bibinfo {pages} {527} (\bibinfo {year} {2023})}\BibitemShut {NoStop}%
\bibitem [{\citenamefont {Guo}\ \emph {et~al.}(2018)\citenamefont {Guo} \emph {et~al.}}]{guo2018evidence}%
  \BibitemOpen
  \bibfield  {author} {\bibinfo {author} {\bibfnamefont {C.}~\bibnamefont {Guo}} \emph {et~al.},\ }\bibfield  {title} {\bibinfo {title} {Evidence for weyl fermions in a canonical heavy-fermion semimetal {YbPtBi}},\ }\href {https://www.nature.com/articles/s41467-018-06782-1} {\bibfield  {journal} {\bibinfo  {journal} {Nat. Commun.}\ }\textbf {\bibinfo {volume} {9}},\ \bibinfo {pages} {4622} (\bibinfo {year} {2018})}\BibitemShut {NoStop}%
\bibitem [{\citenamefont {Cheng}\ \emph {et~al.}(2023)\citenamefont {Cheng} \emph {et~al.}}]{cheng2023observation}%
  \BibitemOpen
  \bibfield  {author} {\bibinfo {author} {\bibfnamefont {Z.-J.}\ \bibnamefont {Cheng}} \emph {et~al.},\ }\bibfield  {title} {\bibinfo {title} {Observation of {K}ondo lattice and {K}ondo-enhanced anomalous {H}all effect in an itinerant ferromagnet},\ }\href {https://arxiv.org/abs/2302.12113} {\bibfield  {journal} {\bibinfo  {journal} {arXiv preprint arXiv:2302.12113}\ } (\bibinfo {year} {2023})}\BibitemShut {NoStop}%
\bibitem [{\citenamefont {Dzsaber}\ \emph {et~al.}(2017)\citenamefont {Dzsaber} \emph {et~al.}}]{dzsaber2017kondo}%
  \BibitemOpen
  \bibfield  {author} {\bibinfo {author} {\bibfnamefont {S.}~\bibnamefont {Dzsaber}} \emph {et~al.},\ }\bibfield  {title} {\bibinfo {title} {Kondo {I}nsulator to {S}emimetal {T}ransformation {T}uned by {S}pin-{O}rbit {C}oupling},\ }\href {https://journals.aps.org/prl/abstract/10.1103/PhysRevLett.118.246601} {\bibfield  {journal} {\bibinfo  {journal} {Phys. Rev. Lett.}\ }\textbf {\bibinfo {volume} {118}},\ \bibinfo {pages} {246601} (\bibinfo {year} {2017})}\BibitemShut {NoStop}%
\bibitem [{\citenamefont {Lai}\ \emph {et~al.}(2018)\citenamefont {Lai}, \citenamefont {Grefe}, \citenamefont {Paschen},\ and\ \citenamefont {Si}}]{lai2018weyl}%
  \BibitemOpen
  \bibfield  {author} {\bibinfo {author} {\bibfnamefont {H.-H.}\ \bibnamefont {Lai}}, \bibinfo {author} {\bibfnamefont {S.~E.}\ \bibnamefont {Grefe}}, \bibinfo {author} {\bibfnamefont {S.}~\bibnamefont {Paschen}},\ and\ \bibinfo {author} {\bibfnamefont {Q.}~\bibnamefont {Si}},\ }\bibfield  {title} {\bibinfo {title} {Weyl--{K}ondo semimetal in heavy-fermion systems},\ }\href {https://www.pnas.org/doi/abs/10.1073/pnas.1715851115} {\bibfield  {journal} {\bibinfo  {journal} {Proc. Natl. Acad. Sci.}\ }\textbf {\bibinfo {volume} {115}},\ \bibinfo {pages} {93} (\bibinfo {year} {2018})}\BibitemShut {NoStop}%
\bibitem [{\citenamefont {Dzsaber}\ \emph {et~al.}(2021)\citenamefont {Dzsaber} \emph {et~al.}}]{dzsaber2021giant}%
  \BibitemOpen
  \bibfield  {author} {\bibinfo {author} {\bibfnamefont {S.}~\bibnamefont {Dzsaber}} \emph {et~al.},\ }\bibfield  {title} {\bibinfo {title} {Giant spontaneous {H}all effect in a nonmagnetic {W}eyl--{K}ondo semimetal},\ }\href {https://www.pnas.org/doi/abs/10.1073/pnas.2013386118} {\bibfield  {journal} {\bibinfo  {journal} {Proc. Natl. Acad. Sci.}\ }\textbf {\bibinfo {volume} {118}},\ \bibinfo {pages} {e2013386118} (\bibinfo {year} {2021})}\BibitemShut {NoStop}%
\bibitem [{\citenamefont {Grefe}\ \emph {et~al.}(2020)\citenamefont {Grefe}, \citenamefont {Lai}, \citenamefont {Paschen},\ and\ \citenamefont {Si}}]{grefe2020weyl}%
  \BibitemOpen
  \bibfield  {author} {\bibinfo {author} {\bibfnamefont {S.~E.}\ \bibnamefont {Grefe}}, \bibinfo {author} {\bibfnamefont {H.-H.}\ \bibnamefont {Lai}}, \bibinfo {author} {\bibfnamefont {S.}~\bibnamefont {Paschen}},\ and\ \bibinfo {author} {\bibfnamefont {Q.}~\bibnamefont {Si}},\ }\bibfield  {title} {\bibinfo {title} {Weyl-{K}ondo semimetals in nonsymmorphic systems},\ }\href {https://journals.aps.org/prb/abstract/10.1103/PhysRevB.101.075138} {\bibfield  {journal} {\bibinfo  {journal} {Phys. Rev. B}\ }\textbf {\bibinfo {volume} {101}},\ \bibinfo {pages} {075138} (\bibinfo {year} {2020})}\BibitemShut {NoStop}%
\bibitem [{\citenamefont {Xie}\ \emph {et~al.}(2021)\citenamefont {Xie}, \citenamefont {Gao}, \citenamefont {Xu}, \citenamefont {Zhang}, \citenamefont {Hu}, \citenamefont {Gao},\ and\ \citenamefont {Law}}]{xie2021kramers}%
  \BibitemOpen
  \bibfield  {author} {\bibinfo {author} {\bibfnamefont {Y.-M.}\ \bibnamefont {Xie}}, \bibinfo {author} {\bibfnamefont {X.-J.}\ \bibnamefont {Gao}}, \bibinfo {author} {\bibfnamefont {X.~Y.}\ \bibnamefont {Xu}}, \bibinfo {author} {\bibfnamefont {C.-P.}\ \bibnamefont {Zhang}}, \bibinfo {author} {\bibfnamefont {J.-X.}\ \bibnamefont {Hu}}, \bibinfo {author} {\bibfnamefont {J.~Z.}\ \bibnamefont {Gao}},\ and\ \bibinfo {author} {\bibfnamefont {K.~T.}\ \bibnamefont {Law}},\ }\bibfield  {title} {\bibinfo {title} {Kramers nodal line metals},\ }\href {https://www.nature.com/articles/s41467-021-22903-9} {\bibfield  {journal} {\bibinfo  {journal} {Nat. Commun.}\ }\textbf {\bibinfo {volume} {12}},\ \bibinfo {pages} {3064} (\bibinfo {year} {2021})}\BibitemShut {NoStop}%
\bibitem [{\citenamefont {Kurumaji}\ \emph {et~al.}(2025)\citenamefont {Kurumaji} \emph {et~al.}}]{kurumaji2025electronic}%
  \BibitemOpen
  \bibfield  {author} {\bibinfo {author} {\bibfnamefont {T.}~\bibnamefont {Kurumaji}} \emph {et~al.},\ }\bibfield  {title} {\bibinfo {title} {Electronic {S}tructure of {K}ramers {N}odal-{L}ine {S}emimetal {YAuGe} and {A}nomalous {H}all {E}ffect {I}nduced by {M}agnetic {R}are-{E}arth {S}ubstitution},\ }\href {https://advanced.onlinelibrary.wiley.com/doi/full/10.1002/advs.202501669} {\bibfield  {journal} {\bibinfo  {journal} {Adv. Sci.}\ }\textbf {\bibinfo {volume} {12}},\ \bibinfo {pages} {2501669} (\bibinfo {year} {2025})}\BibitemShut {NoStop}%
\bibitem [{\citenamefont {Zhang}\ \emph {et~al.}(2023)\citenamefont {Zhang} \emph {et~al.}}]{zhang2023kramers}%
  \BibitemOpen
  \bibfield  {author} {\bibinfo {author} {\bibfnamefont {Y.}~\bibnamefont {Zhang}} \emph {et~al.},\ }\bibfield  {title} {\bibinfo {title} {Kramers nodal lines and {W}eyl fermions in {SmAlSi}},\ }\href {https://www.nature.com/articles/s42005-023-01257-2} {\bibfield  {journal} {\bibinfo  {journal} {Comms. Phys.}\ }\textbf {\bibinfo {volume} {6}},\ \bibinfo {pages} {134} (\bibinfo {year} {2023})}\BibitemShut {NoStop}%
\bibitem [{\citenamefont {Zhang}\ \emph {et~al.}(2025)\citenamefont {Zhang} \emph {et~al.}}]{zhang2025kramers}%
  \BibitemOpen
  \bibfield  {author} {\bibinfo {author} {\bibfnamefont {Y.}~\bibnamefont {Zhang}} \emph {et~al.},\ }\bibfield  {title} {\bibinfo {title} {Kramers nodal lines in intercalated {TaS$_{2}$} superconductors},\ }\href {https://www.nature.com/articles/s41467-025-60020-z} {\bibfield  {journal} {\bibinfo  {journal} {Nat. Commun.}\ }\textbf {\bibinfo {volume} {16}},\ \bibinfo {pages} {4984} (\bibinfo {year} {2025})}\BibitemShut {NoStop}%
\bibitem [{\citenamefont {Sarkar}\ \emph {et~al.}(2023)\citenamefont {Sarkar} \emph {et~al.}}]{sarkar2023charge}%
  \BibitemOpen
  \bibfield  {author} {\bibinfo {author} {\bibfnamefont {S.}~\bibnamefont {Sarkar}} \emph {et~al.},\ }\bibfield  {title} {\bibinfo {title} {Charge density wave induced nodal lines in {LaTe$_3$}},\ }\href {https://www.nature.com/articles/s41467-023-39271-1} {\bibfield  {journal} {\bibinfo  {journal} {Nat. Commun.}\ }\textbf {\bibinfo {volume} {14}},\ \bibinfo {pages} {3628} (\bibinfo {year} {2023})}\BibitemShut {NoStop}%
\bibitem [{\citenamefont {Shang}\ \emph {et~al.}(2022)\citenamefont {Shang} \emph {et~al.}}]{shang2022unconventional}%
  \BibitemOpen
  \bibfield  {author} {\bibinfo {author} {\bibfnamefont {T.}~\bibnamefont {Shang}} \emph {et~al.},\ }\bibfield  {title} {\bibinfo {title} {Unconventional superconductivity in topological {K}ramers nodal-line semimetals},\ }\href {https://www.science.org/doi/full/10.1126/sciadv.abq6589} {\bibfield  {journal} {\bibinfo  {journal} {Sci. Adv.}\ }\textbf {\bibinfo {volume} {8}},\ \bibinfo {pages} {eabq6589} (\bibinfo {year} {2022})}\BibitemShut {NoStop}%
\bibitem [{\citenamefont {Sarkar}\ \emph {et~al.}(2026)\citenamefont {Sarkar} \emph {et~al.}}]{sarkar2026kramers}%
  \BibitemOpen
  \bibfield  {author} {\bibinfo {author} {\bibfnamefont {S.}~\bibnamefont {Sarkar}} \emph {et~al.},\ }\bibfield  {title} {\bibinfo {title} {Kramers nodal line in the charge density wave state of {YTe$_3$} and the influence of twin domains},\ }\href {https://journals.aps.org/prb/abstract/10.1103/ld4p-hl13} {\bibfield  {journal} {\bibinfo  {journal} {Phys. Rev. B.}\ }\textbf {\bibinfo {volume} {113}},\ \bibinfo {pages} {035129} (\bibinfo {year} {2026})}\BibitemShut {NoStop}%
\bibitem [{\citenamefont {Chen}\ \emph {et~al.}(2022)\citenamefont {Chen} \emph {et~al.}}]{chen2022topological}%
  \BibitemOpen
  \bibfield  {author} {\bibinfo {author} {\bibfnamefont {L.}~\bibnamefont {Chen}} \emph {et~al.},\ }\bibfield  {title} {\bibinfo {title} {Topological semimetal driven by strong correlations and crystalline symmetry},\ }\href {https://www.nature.com/articles/s41567-022-01743-4} {\bibfield  {journal} {\bibinfo  {journal} {Nat. Phys.}\ }\textbf {\bibinfo {volume} {18}},\ \bibinfo {pages} {1341} (\bibinfo {year} {2022})}\BibitemShut {NoStop}%
\bibitem [{\citenamefont {Hirschmann}\ \emph {et~al.}(2021)\citenamefont {Hirschmann}, \citenamefont {Leonhardt}, \citenamefont {Kilic}, \citenamefont {Fabini},\ and\ \citenamefont {Schnyder}}]{hirschmann2021symmetry}%
  \BibitemOpen
  \bibfield  {author} {\bibinfo {author} {\bibfnamefont {M.~M.}\ \bibnamefont {Hirschmann}}, \bibinfo {author} {\bibfnamefont {A.}~\bibnamefont {Leonhardt}}, \bibinfo {author} {\bibfnamefont {B.}~\bibnamefont {Kilic}}, \bibinfo {author} {\bibfnamefont {D.~H.}\ \bibnamefont {Fabini}},\ and\ \bibinfo {author} {\bibfnamefont {A.~P.}\ \bibnamefont {Schnyder}},\ }\bibfield  {title} {\bibinfo {title} {Symmetry-enforced band crossings in tetragonal materials: {D}irac and {W}eyl degeneracies on points, lines, and planes},\ }\href {https://journals.aps.org/prmaterials/abstract/10.1103/PhysRevMaterials.5.054202} {\bibfield  {journal} {\bibinfo  {journal} {Phys. Rev. Mater.}\ }\textbf {\bibinfo {volume} {5}},\ \bibinfo {pages} {054202} (\bibinfo {year} {2021})}\BibitemShut {NoStop}%
\bibitem [{\citenamefont {Shen}\ \emph {et~al.}(2026)\citenamefont {Shen} \emph {et~al.}}]{shen2026giant}%
  \BibitemOpen
  \bibfield  {author} {\bibinfo {author} {\bibfnamefont {Z.}~\bibnamefont {Shen}} \emph {et~al.},\ }\bibfield  {title} {\bibinfo {title} {Giant field induced unconventional anomalous {H}all conductivity in the noncentrosymmetric {W}eyl metal {CeCoGe$_{3}$}},\ }\href {https://journals.aps.org/prb/abstract/10.1103/5psl-z6xm} {\bibfield  {journal} {\bibinfo  {journal} {Phys. Rev. B.}\ }\textbf {\bibinfo {volume} {113}},\ \bibinfo {pages} {L081107} (\bibinfo {year} {2026})}\BibitemShut {NoStop}%
\bibitem [{\citenamefont {Cao}\ \emph {et~al.}(2020)\citenamefont {Cao}, \citenamefont {Zhi},\ and\ \citenamefont {Zhu}}]{cao2020trivial}%
  \BibitemOpen
  \bibfield  {author} {\bibinfo {author} {\bibfnamefont {C.}~\bibnamefont {Cao}}, \bibinfo {author} {\bibfnamefont {G.-X.}\ \bibnamefont {Zhi}},\ and\ \bibinfo {author} {\bibfnamefont {J.-X.}\ \bibnamefont {Zhu}},\ }\bibfield  {title} {\bibinfo {title} {From {T}rivial {K}ondo {I}nsulator {Ce$_3$Pt$_3$Bi$_4$} to {T}opological {N}odal-{L}ine {S}emimetal {Ce$_3$Pd$_3$Bi$_4$}},\ }\href {https://journals.aps.org/prl/abstract/10.1103/PhysRevLett.124.166403} {\bibfield  {journal} {\bibinfo  {journal} {Phys. Rev. Lett.}\ }\textbf {\bibinfo {volume} {124}},\ \bibinfo {pages} {166403} (\bibinfo {year} {2020})}\BibitemShut {NoStop}%
\bibitem [{\citenamefont {Li}\ \emph {et~al.}(2025)\citenamefont {Li}, \citenamefont {Li}, \citenamefont {Yin}, \citenamefont {Yang},\ and\ \citenamefont {Jia}}]{li2025extrinsic}%
  \BibitemOpen
  \bibfield  {author} {\bibinfo {author} {\bibfnamefont {L.}~\bibnamefont {Li}}, \bibinfo {author} {\bibfnamefont {J.}~\bibnamefont {Li}}, \bibinfo {author} {\bibfnamefont {J.-X.}\ \bibnamefont {Yin}}, \bibinfo {author} {\bibfnamefont {Y.-f.}\ \bibnamefont {Yang}},\ and\ \bibinfo {author} {\bibfnamefont {S.}~\bibnamefont {Jia}},\ }\bibfield  {title} {\bibinfo {title} {Extrinsic anomalous {H}all effect in four cerium-based heavy fermion ferromagnets},\ }\href {https://journals.aps.org/prb/abstract/10.1103/jl69-mpvr} {\bibfield  {journal} {\bibinfo  {journal} {Phys. Rev. B.}\ }\textbf {\bibinfo {volume} {112}},\ \bibinfo {pages} {165142} (\bibinfo {year} {2025})}\BibitemShut {NoStop}%
\bibitem [{\citenamefont {Xu}\ \emph {et~al.}(2024)\citenamefont {Xu} \emph {et~al.}}]{xu2024tunable}%
  \BibitemOpen
  \bibfield  {author} {\bibinfo {author} {\bibfnamefont {Q.}~\bibnamefont {Xu}} \emph {et~al.},\ }\bibfield  {title} {\bibinfo {title} {Tunable giant anomalous {H}all effect in the {K}ondo-lattice ferromagnet {UBiTe}},\ }\href {https://journals.aps.org/prb/abstract/10.1103/PhysRevB.110.165162} {\bibfield  {journal} {\bibinfo  {journal} {Phys. Rev. B.}\ }\textbf {\bibinfo {volume} {110}},\ \bibinfo {pages} {165162} (\bibinfo {year} {2024})}\BibitemShut {NoStop}%
\bibitem [{\citenamefont {Klicpera}\ and\ \citenamefont {Javorsk{\`y}}(2014)}]{klicpera2014study}%
  \BibitemOpen
  \bibfield  {author} {\bibinfo {author} {\bibfnamefont {M.}~\bibnamefont {Klicpera}}\ and\ \bibinfo {author} {\bibfnamefont {P.}~\bibnamefont {Javorsk{\`y}}},\ }\bibfield  {title} {\bibinfo {title} {Study of electronic properties in {$RCu_{1-x}Au_{x}Al_{3}$} compounds, where {R= Ce, La}},\ }\href {https://www.sciencedirect.com/science/article/pii/S0304885314003011} {\bibfield  {journal} {\bibinfo  {journal} {J. Magn. Magn. Mater.}\ }\textbf {\bibinfo {volume} {363}},\ \bibinfo {pages} {88} (\bibinfo {year} {2014})}\BibitemShut {NoStop}%
\bibitem [{\citenamefont {Aoki}\ \emph {et~al.}(2000)\citenamefont {Aoki}, \citenamefont {Saha}, \citenamefont {Matsuda}, \citenamefont {Sugawara},\ and\ \citenamefont {Sato}}]{aoki2000anomalous}%
  \BibitemOpen
  \bibfield  {author} {\bibinfo {author} {\bibfnamefont {Y.}~\bibnamefont {Aoki}}, \bibinfo {author} {\bibfnamefont {S.~R.}\ \bibnamefont {Saha}}, \bibinfo {author} {\bibfnamefont {T.~D.}\ \bibnamefont {Matsuda}}, \bibinfo {author} {\bibfnamefont {H.}~\bibnamefont {Sugawara}},\ and\ \bibinfo {author} {\bibfnamefont {H.}~\bibnamefont {Sato}},\ }\bibfield  {title} {\bibinfo {title} {Anomalous low-energy excitation in {{CeAuAl}\(_{3}\)}},\ }\href {https://doi.org/10.1016/S0921-4526(99)01006-6} {\bibfield  {journal} {\bibinfo  {journal} {Physica B}\ }\textbf {\bibinfo {volume} {281--282}},\ \bibinfo {pages} {110} (\bibinfo {year} {2000})}\BibitemShut {NoStop}%
\bibitem [{\citenamefont {{\v{C}}erm{\'a}k}\ \emph {et~al.}(2019)\citenamefont {{\v{C}}erm{\'a}k}, \citenamefont {Schneidewind}, \citenamefont {Liu}, \citenamefont {Koza}, \citenamefont {Franz}, \citenamefont {Sch{\"o}nmann}, \citenamefont {Sobolev},\ and\ \citenamefont {Pfleiderer}}]{vcermak2019magnetoelastic}%
  \BibitemOpen
  \bibfield  {author} {\bibinfo {author} {\bibfnamefont {P.}~\bibnamefont {{\v{C}}erm{\'a}k}}, \bibinfo {author} {\bibfnamefont {A.}~\bibnamefont {Schneidewind}}, \bibinfo {author} {\bibfnamefont {B.}~\bibnamefont {Liu}}, \bibinfo {author} {\bibfnamefont {M.~M.}\ \bibnamefont {Koza}}, \bibinfo {author} {\bibfnamefont {C.}~\bibnamefont {Franz}}, \bibinfo {author} {\bibfnamefont {R.}~\bibnamefont {Sch{\"o}nmann}}, \bibinfo {author} {\bibfnamefont {O.}~\bibnamefont {Sobolev}},\ and\ \bibinfo {author} {\bibfnamefont {C.}~\bibnamefont {Pfleiderer}},\ }\bibfield  {title} {\bibinfo {title} {Magnetoelastic hybrid excitations in {CeAuAl$_3$}},\ }\href {https://www.pnas.org/doi/abs/10.1073/pnas.1819664116} {\bibfield  {journal} {\bibinfo  {journal} {Proc. Natl. Acad. Sci.}\ }\textbf {\bibinfo {volume} {116}},\ \bibinfo {pages} {6695} (\bibinfo {year} {2019})}\BibitemShut {NoStop}%
\bibitem [{\citenamefont {Adroja}\ \emph {et~al.}(2015)\citenamefont {Adroja} \emph {et~al.}}]{adroja2015muon}%
  \BibitemOpen
  \bibfield  {author} {\bibinfo {author} {\bibfnamefont {D.}~\bibnamefont {Adroja}} \emph {et~al.},\ }\bibfield  {title} {\bibinfo {title} {Muon spin rotation and neutron scattering study of the noncentrosymmetric tetragonal compound {CeAuAl$_{3}$}},\ }\href {https://journals.aps.org/prb/abstract/10.1103/PhysRevB.91.134425} {\bibfield  {journal} {\bibinfo  {journal} {Phys. Rev. B}\ }\textbf {\bibinfo {volume} {91}},\ \bibinfo {pages} {134425} (\bibinfo {year} {2015})}\BibitemShut {NoStop}%
\bibitem [{\citenamefont {Sugawara}\ \emph {et~al.}(1999)\citenamefont {Sugawara} \emph {et~al.}}]{sugawara1999magnetic}%
  \BibitemOpen
  \bibfield  {author} {\bibinfo {author} {\bibfnamefont {H.}~\bibnamefont {Sugawara}} \emph {et~al.},\ }\bibfield  {title} {\bibinfo {title} {Magnetic and transport properties in {CeAuAl$_{3}$} single crystal},\ }\href {https://www.sciencedirect.com/science/article/pii/S092145269800814X} {\bibfield  {journal} {\bibinfo  {journal} {Physica B: Condensed Matter}\ }\textbf {\bibinfo {volume} {259}},\ \bibinfo {pages} {16} (\bibinfo {year} {1999})}\BibitemShut {NoStop}%
\bibitem [{\citenamefont {Ullah}\ \emph {et~al.}(2024)\citenamefont {Ullah}, \citenamefont {Ali}, \citenamefont {Murad},\ and\ \citenamefont {Mehmood}}]{ullah2024electronic}%
  \BibitemOpen
  \bibfield  {author} {\bibinfo {author} {\bibfnamefont {I.}~\bibnamefont {Ullah}}, \bibinfo {author} {\bibfnamefont {Z.}~\bibnamefont {Ali}}, \bibinfo {author} {\bibfnamefont {M.}~\bibnamefont {Murad}},\ and\ \bibinfo {author} {\bibfnamefont {S.}~\bibnamefont {Mehmood}},\ }\bibfield  {title} {\bibinfo {title} {Electronic {S}tructure, {L}ong {R}ange {M}agnetic {O}rder and {E}lastic {P}roperties of {C}erium {B}ased {N}on-centro {S}ymmetric {I}ntermetallics {CeTAl$_{3}$} {(T= Pd, Pt, Cu, Ag and Au)}},\ }\href {https://link.springer.com/article/10.1007/s10904-024-03023-5} {\bibfield  {journal} {\bibinfo  {journal} {J. Inorg. Organomet. Polym. Mater.}\ }\textbf {\bibinfo {volume} {34}},\ \bibinfo {pages} {3148} (\bibinfo {year} {2024})}\BibitemShut {NoStop}%
\bibitem [{\citenamefont {Paschen}\ \emph {et~al.}(1998)\citenamefont {Paschen}, \citenamefont {Felder},\ and\ \citenamefont {Ott}}]{paschen1998transport}%
  \BibitemOpen
  \bibfield  {author} {\bibinfo {author} {\bibfnamefont {S.}~\bibnamefont {Paschen}}, \bibinfo {author} {\bibfnamefont {E.}~\bibnamefont {Felder}},\ and\ \bibinfo {author} {\bibfnamefont {H.}~\bibnamefont {Ott}},\ }\bibfield  {title} {\bibinfo {title} {Transport and thermodynamic properties of {CeAuAl$_3$}},\ }\href {https://link.springer.com/article/10.1007/s100510050237} {\bibfield  {journal} {\bibinfo  {journal} {Eur. Phys. J. B}\ }\textbf {\bibinfo {volume} {2}},\ \bibinfo {pages} {169} (\bibinfo {year} {1998})}\BibitemShut {NoStop}%
\bibitem [{\citenamefont {Vonlanthen}\ \emph {et~al.}(1999)\citenamefont {Vonlanthen}, \citenamefont {Gavilano}, \citenamefont {Ambrosini},\ and\ \citenamefont {Ott}}]{vonlanthen199927al}%
  \BibitemOpen
  \bibfield  {author} {\bibinfo {author} {\bibfnamefont {P.}~\bibnamefont {Vonlanthen}}, \bibinfo {author} {\bibfnamefont {J.}~\bibnamefont {Gavilano}}, \bibinfo {author} {\bibfnamefont {B.}~\bibnamefont {Ambrosini}},\ and\ \bibinfo {author} {\bibfnamefont {H.}~\bibnamefont {Ott}},\ }\bibfield  {title} {\bibinfo {title} {$^{27}${Al} nuclear magnetic resonance studies of {CeAuAl$_{3}$}},\ }\href {https://www.sciencedirect.com/science/article/pii/S0921452698010394} {\bibfield  {journal} {\bibinfo  {journal} {Phys. B Condens. Matter.}\ }\textbf {\bibinfo {volume} {259}},\ \bibinfo {pages} {18} (\bibinfo {year} {1999})}\BibitemShut {NoStop}%
\bibitem [{\citenamefont {Deyerling}\ \emph {et~al.}(2023)\citenamefont {Deyerling}, \citenamefont {Wilde},\ and\ \citenamefont {Pfleiderer}}]{deyerling2023electronic}%
  \BibitemOpen
  \bibfield  {author} {\bibinfo {author} {\bibfnamefont {A.}~\bibnamefont {Deyerling}}, \bibinfo {author} {\bibfnamefont {M.~A.}\ \bibnamefont {Wilde}},\ and\ \bibinfo {author} {\bibfnamefont {C.}~\bibnamefont {Pfleiderer}},\ }\bibfield  {title} {\bibinfo {title} {Electronic structure of {CeAuAl$_{3}$} using density functional theory},\ }\href {https://www.scipost.org/SciPostPhysProc.11.008?acad_field_slug=physics} {\bibfield  {journal} {\bibinfo  {journal} {SciPost Phys. Proc.}\ ,\ \bibinfo {pages} {008}} (\bibinfo {year} {2023})}\BibitemShut {NoStop}%
\bibitem [{\citenamefont {Franz}\ \emph {et~al.}(2016)\citenamefont {Franz} \emph {et~al.}}]{franz2016single}%
  \BibitemOpen
  \bibfield  {author} {\bibinfo {author} {\bibfnamefont {C.}~\bibnamefont {Franz}} \emph {et~al.},\ }\bibfield  {title} {\bibinfo {title} {Single crystal growth of {CeTAl$_{3}$} {(T= Cu, Ag, Au, Pd and Pt)}},\ }\href {https://www.sciencedirect.com/science/article/pii/S0925838816321041} {\bibfield  {journal} {\bibinfo  {journal} {J. Alloys Compd.}\ }\textbf {\bibinfo {volume} {688}},\ \bibinfo {pages} {978} (\bibinfo {year} {2016})}\BibitemShut {NoStop}%
\bibitem [{\citenamefont {Mock}\ \emph {et~al.}(1999)\citenamefont {Mock}, \citenamefont {Pfleiderer},\ and\ \citenamefont {L{\"o}hneysen}}]{mock1999low}%
  \BibitemOpen
  \bibfield  {author} {\bibinfo {author} {\bibfnamefont {S.}~\bibnamefont {Mock}}, \bibinfo {author} {\bibfnamefont {C.}~\bibnamefont {Pfleiderer}},\ and\ \bibinfo {author} {\bibfnamefont {H.~v.}\ \bibnamefont {L{\"o}hneysen}},\ }\bibfield  {title} {\bibinfo {title} {Low-{T}emperature {P}roperties of {CeTAl$_{3}$} {(T= Au,Cu,Pt)} and {CeAuGa$_{3}$}},\ }\href {https://link.springer.com/article/10.1023/a:1021838628126} {\bibfield  {journal} {\bibinfo  {journal} {J. Low. Temp. Phys.}\ }\textbf {\bibinfo {volume} {115}},\ \bibinfo {pages} {1} (\bibinfo {year} {1999})}\BibitemShut {NoStop}%
\bibitem [{\citenamefont {Liu}\ \emph {et~al.}(2018)\citenamefont {Liu}, \citenamefont {{\v{C}}erm{\'a}k}, \citenamefont {Franz}, \citenamefont {Pfleiderer},\ and\ \citenamefont {Schneidewind}}]{liu2018lattice}%
  \BibitemOpen
  \bibfield  {author} {\bibinfo {author} {\bibfnamefont {B.-Q.}\ \bibnamefont {Liu}}, \bibinfo {author} {\bibfnamefont {P.}~\bibnamefont {{\v{C}}erm{\'a}k}}, \bibinfo {author} {\bibfnamefont {C.}~\bibnamefont {Franz}}, \bibinfo {author} {\bibfnamefont {C.}~\bibnamefont {Pfleiderer}},\ and\ \bibinfo {author} {\bibfnamefont {A.}~\bibnamefont {Schneidewind}},\ }\bibfield  {title} {\bibinfo {title} {Lattice dynamics and coupled quadrupole-phonon excitations in {CeAuAl$_{3}$}},\ }\href {https://journals.aps.org/prb/abstract/10.1103/PhysRevB.98.174306} {\bibfield  {journal} {\bibinfo  {journal} {Phys. Rev. B.}\ }\textbf {\bibinfo {volume} {98}},\ \bibinfo {pages} {174306} (\bibinfo {year} {2018})}\BibitemShut {NoStop}%
\bibitem [{\citenamefont {Elcoro}\ \emph {et~al.}(2017)\citenamefont {Elcoro} \emph {et~al.}}]{elcoro2017double}%
  \BibitemOpen
  \bibfield  {author} {\bibinfo {author} {\bibfnamefont {L.}~\bibnamefont {Elcoro}} \emph {et~al.},\ }\bibfield  {title} {\bibinfo {title} {Double crystallographic groups and their representations on the {B}ilbao {C}rystallographic {S}erver},\ }\href {https://onlinelibrary.wiley.com/iucr/doi/10.1107/S1600576717011712} {\bibfield  {journal} {\bibinfo  {journal} {Applied Crystallography}\ }\textbf {\bibinfo {volume} {50}},\ \bibinfo {pages} {1457} (\bibinfo {year} {2017})}\BibitemShut {NoStop}%
\bibitem [{\citenamefont {Desgranges}\ and\ \citenamefont {Schotte}(1982)}]{desgranges1982specific}%
  \BibitemOpen
  \bibfield  {author} {\bibinfo {author} {\bibfnamefont {H.-U.}\ \bibnamefont {Desgranges}}\ and\ \bibinfo {author} {\bibfnamefont {K.}~\bibnamefont {Schotte}},\ }\bibfield  {title} {\bibinfo {title} {Specific heat of the {K}ondo model},\ }\href {https://www.sciencedirect.com/science/article/pii/0375960182904819} {\bibfield  {journal} {\bibinfo  {journal} {Phys. Lett. A.}\ }\textbf {\bibinfo {volume} {91}},\ \bibinfo {pages} {240} (\bibinfo {year} {1982})}\BibitemShut {NoStop}%
\bibitem [{\citenamefont {Yang}\ \emph {et~al.}(2008)\citenamefont {Yang}, \citenamefont {Fisk}, \citenamefont {Lee}, \citenamefont {Thompson},\ and\ \citenamefont {Pines}}]{yang2008scaling}%
  \BibitemOpen
  \bibfield  {author} {\bibinfo {author} {\bibfnamefont {Y.-f.}\ \bibnamefont {Yang}}, \bibinfo {author} {\bibfnamefont {Z.}~\bibnamefont {Fisk}}, \bibinfo {author} {\bibfnamefont {H.-O.}\ \bibnamefont {Lee}}, \bibinfo {author} {\bibfnamefont {J.}~\bibnamefont {Thompson}},\ and\ \bibinfo {author} {\bibfnamefont {D.}~\bibnamefont {Pines}},\ }\bibfield  {title} {\bibinfo {title} {Scaling the {K}ondo lattice},\ }\href {https://www.nature.com/articles/nature07157} {\bibfield  {journal} {\bibinfo  {journal} {Nature}\ }\textbf {\bibinfo {volume} {454}},\ \bibinfo {pages} {611} (\bibinfo {year} {2008})}\BibitemShut {NoStop}%
\bibitem [{\citenamefont {Arushi}\ \emph {et~al.}(2026)\citenamefont {Arushi} \emph {et~al.}}]{allen2026chiral}%
  \BibitemOpen
  \bibfield  {author} {\bibinfo {author} {\bibnamefont {Arushi}} \emph {et~al.},\ }\bibfield  {title} {\bibinfo {title} {Chiral {W}eyl-{K}ondo semimetallic state through enhanced correlation in {CeGaGe}},\ }\href {https://arxiv.org/pdf/2609.32012} {\bibfield  {journal} {\bibinfo  {journal} {arXiv preprint arXiv:2609.32012}\ } (\bibinfo {year} {2026})}\BibitemShut {NoStop}%
\bibitem [{\citenamefont {Adroja}\ \emph {et~al.}(2004)\citenamefont {Adroja} \emph {et~al.}}]{adroja2004reduced}%
  \BibitemOpen
  \bibfield  {author} {\bibinfo {author} {\bibfnamefont {D.}~\bibnamefont {Adroja}} \emph {et~al.},\ }\bibfield  {title} {\bibinfo {title} {A reduced moment antiferromagnetic {K}ondo lattice compound: {Ce$_{8}$Pd$_{24}$Ga}},\ }\href {https://www.sciencedirect.com/science/article/pii/S0304885303026350} {\bibfield  {journal} {\bibinfo  {journal} {J. Magn. Magn. Mater.}\ }\textbf {\bibinfo {volume} {272}},\ \bibinfo {pages} {622} (\bibinfo {year} {2004})}\BibitemShut {NoStop}%
\bibitem [{\citenamefont {Ueland}\ \emph {et~al.}(2018)\citenamefont {Ueland} \emph {et~al.}}]{ueland2018reduction}%
  \BibitemOpen
  \bibfield  {author} {\bibinfo {author} {\bibfnamefont {B.}~\bibnamefont {Ueland}} \emph {et~al.},\ }\bibfield  {title} {\bibinfo {title} {Reduction of the ordered magnetic moment and its relationship to {K}ondo coherence in {Ce$_{1- x}$La$_{x}$Cu$_{2}$Ge$_{2}$}},\ }\href {https://journals.aps.org/prb/abstract/10.1103/PhysRevB.97.165121} {\bibfield  {journal} {\bibinfo  {journal} {Phys. Rev. B.}\ }\textbf {\bibinfo {volume} {97}},\ \bibinfo {pages} {165121} (\bibinfo {year} {2018})}\BibitemShut {NoStop}%
\bibitem [{\citenamefont {Kratochv{\'\i}lov{\'a}}\ \emph {et~al.}(2015)\citenamefont {Kratochv{\'\i}lov{\'a}} \emph {et~al.}}]{kratochvilova2015coexistence}%
  \BibitemOpen
  \bibfield  {author} {\bibinfo {author} {\bibfnamefont {M.}~\bibnamefont {Kratochv{\'\i}lov{\'a}}} \emph {et~al.},\ }\bibfield  {title} {\bibinfo {title} {Coexistence of {A}ntiferromagnetism and {S}uperconductivity in {H}eavy {F}ermion {C}erium {C}ompound {Ce$_{3}$PdIn$_{11}$}},\ }\href {https://www.nature.com/articles/srep15904} {\bibfield  {journal} {\bibinfo  {journal} {Sci. Rep.}\ }\textbf {\bibinfo {volume} {5}},\ \bibinfo {pages} {15904} (\bibinfo {year} {2015})}\BibitemShut {NoStop}%
\bibitem [{\citenamefont {Sluchanko}\ \emph {et~al.}(2015)\citenamefont {Sluchanko} \emph {et~al.}}]{sluchanko2015anomalies}%
  \BibitemOpen
  \bibfield  {author} {\bibinfo {author} {\bibfnamefont {N.}~\bibnamefont {Sluchanko}} \emph {et~al.},\ }\bibfield  {title} {\bibinfo {title} {Anomalies of magnetoresistance in {Ce}-based heavy fermion compounds},\ }\href {https://pubs.aip.org/aip/ltp/article/41/12/1011/251861} {\bibfield  {journal} {\bibinfo  {journal} {Low Temp. Phys.}\ }\textbf {\bibinfo {volume} {41}},\ \bibinfo {pages} {1011} (\bibinfo {year} {2015})}\BibitemShut {NoStop}%
\bibitem [{\citenamefont {Kondo}(1964)}]{kondo1964resistance}%
  \BibitemOpen
  \bibfield  {author} {\bibinfo {author} {\bibfnamefont {J.}~\bibnamefont {Kondo}},\ }\bibfield  {title} {\bibinfo {title} {Resistance {M}inimum in {D}ilute {M}agnetic {A}lloys},\ }\href {https://academic.oup.com/ptp/article/32/1/37/1834632} {\bibfield  {journal} {\bibinfo  {journal} {Prog. Theor. Phys.}\ }\textbf {\bibinfo {volume} {32}},\ \bibinfo {pages} {37} (\bibinfo {year} {1964})}\BibitemShut {NoStop}%
\bibitem [{\citenamefont {Li}\ \emph {et~al.}(2026)\citenamefont {Li} \emph {et~al.}}]{li2026giant}%
  \BibitemOpen
  \bibfield  {author} {\bibinfo {author} {\bibfnamefont {L.}~\bibnamefont {Li}} \emph {et~al.},\ }\bibfield  {title} {\bibinfo {title} {Giant anomalous {H}all and {N}ernst effects in a heavy fermion ferromagnet},\ }\href {https://www.sciencedirect.com/science/article/pii/S2095927326003907?casa_token=qU_LE-OYx0cAAAAA:b8Wba4dm093GIldQpLnZ20tF6KxYYbV2VYdpuZKJPSq4xldHYCNjaYcdiWV06XaU7JIBNJDUlQ} {\bibfield  {journal} {\bibinfo  {journal} {Sci. Bull.}\ } (\bibinfo {year} {2026})}\BibitemShut {NoStop}%
\bibitem [{\citenamefont {Kotegawa}\ \emph {et~al.}(2024)\citenamefont {Kotegawa} \emph {et~al.}}]{kotegawa2024large}%
  \BibitemOpen
  \bibfield  {author} {\bibinfo {author} {\bibfnamefont {H.}~\bibnamefont {Kotegawa}} \emph {et~al.},\ }\bibfield  {title} {\bibinfo {title} {Large {A}nomalous {H}all {C}onductivity {D}erived from an f-{E}lectron {C}ollinear {A}ntiferromagnetic {S}tructure},\ }\href {https://journals.aps.org/prl/abstract/10.1103/PhysRevLett.133.106301} {\bibfield  {journal} {\bibinfo  {journal} {Phys. Rev. Lett.}\ }\textbf {\bibinfo {volume} {133}},\ \bibinfo {pages} {106301} (\bibinfo {year} {2024})}\BibitemShut {NoStop}%
\bibitem [{\citenamefont {Alam}\ \emph {et~al.}(2023)\citenamefont {Alam} \emph {et~al.}}]{alam2023sign}%
  \BibitemOpen
  \bibfield  {author} {\bibinfo {author} {\bibfnamefont {M.~S.}\ \bibnamefont {Alam}} \emph {et~al.},\ }\bibfield  {title} {\bibinfo {title} {Sign change of anomalous {H}all effect and anomalous {N}ernst effect in the {W}eyl semimetal {CeAlSi}},\ }\href {https://arxiv.org/abs/2210.09764} {\bibfield  {journal} {\bibinfo  {journal} {Phys. Rev. B.}\ }\textbf {\bibinfo {volume} {107}},\ \bibinfo {pages} {085102} (\bibinfo {year} {2023})}\BibitemShut {NoStop}%
\bibitem [{\citenamefont {Kumar}\ \emph {et~al.}(2021)\citenamefont {Kumar} \emph {et~al.}}]{kumar2021giant}%
  \BibitemOpen
  \bibfield  {author} {\bibinfo {author} {\bibfnamefont {N.}~\bibnamefont {Kumar}} \emph {et~al.},\ }\bibfield  {title} {\bibinfo {title} {Giant {A}nomalous {H}all {C}onductivity in the {I}tinerant {F}erromagnet {LaCrSb$_{3}$} and the {E}ffect of f-{E}lectrons},\ }\href {https://advanced.onlinelibrary.wiley.com/doi/full/10.1002/qute.202100023} {\bibfield  {journal} {\bibinfo  {journal} {Adv. Quantum Technol.}\ }\textbf {\bibinfo {volume} {4}},\ \bibinfo {pages} {2100023} (\bibinfo {year} {2021})}\BibitemShut {NoStop}%
\bibitem [{\citenamefont {Allen}\ \emph {et~al.}(2026)\citenamefont {Allen}, \citenamefont {Clements} \emph {et~al.}}]{allen2026large}%
  \BibitemOpen
  \bibfield  {author} {\bibinfo {author} {\bibfnamefont {K.}~\bibnamefont {Allen}}, \bibinfo {author} {\bibfnamefont {E.~M.}\ \bibnamefont {Clements}}, \emph {et~al.},\ }\bibfield  {title} {\bibinfo {title} {Large anomalous {H}all effect in the centrosymmetric kondo lattice {Ce$_3$Al$_{11}$}}} (\bibinfo {year} {2026}),\ \bibinfo {note} {to appear}\BibitemShut {NoStop}%
\bibitem [{\citenamefont {Nagaosa}\ \emph {et~al.}(2010)\citenamefont {Nagaosa}, \citenamefont {Sinova}, \citenamefont {Onoda}, \citenamefont {MacDonald},\ and\ \citenamefont {Ong}}]{nagaosa2010anomalous}%
  \BibitemOpen
  \bibfield  {author} {\bibinfo {author} {\bibfnamefont {N.}~\bibnamefont {Nagaosa}}, \bibinfo {author} {\bibfnamefont {J.}~\bibnamefont {Sinova}}, \bibinfo {author} {\bibfnamefont {S.}~\bibnamefont {Onoda}}, \bibinfo {author} {\bibfnamefont {A.~H.}\ \bibnamefont {MacDonald}},\ and\ \bibinfo {author} {\bibfnamefont {N.~P.}\ \bibnamefont {Ong}},\ }\bibfield  {title} {\bibinfo {title} {Anomalous {H}all effect},\ }\href {https://journals.aps.org/rmp/abstract/10.1103/RevModPhys.82.1539} {\bibfield  {journal} {\bibinfo  {journal} {Rev. Mod. Phys.}\ }\textbf {\bibinfo {volume} {82}},\ \bibinfo {pages} {1539} (\bibinfo {year} {2010})}\BibitemShut {NoStop}%
\bibitem [{\citenamefont {Yang}\ \emph {et~al.}(2025)\citenamefont {Yang} \emph {et~al.}}]{yang2025modulation}%
  \BibitemOpen
  \bibfield  {author} {\bibinfo {author} {\bibfnamefont {J.}~\bibnamefont {Yang}} \emph {et~al.},\ }\bibfield  {title} {\bibinfo {title} {Modulation of the anomalous {H}all angle in a magnetic topological semimetal},\ }\href {https://www.nature.com/articles/s41928-025-01364-8} {\bibfield  {journal} {\bibinfo  {journal} {Nat. Electron.}\ }\textbf {\bibinfo {volume} {8}},\ \bibinfo {pages} {386} (\bibinfo {year} {2025})}\BibitemShut {NoStop}%
\bibitem [{\citenamefont {Berger}(1970)}]{berger1970side}%
  \BibitemOpen
  \bibfield  {author} {\bibinfo {author} {\bibfnamefont {L.}~\bibnamefont {Berger}},\ }\bibfield  {title} {\bibinfo {title} {Side-{J}ump {M}echanism for the {H}all {E}ffect of {F}erromagnets},\ }\href {https://journals.aps.org/prb/abstract/10.1103/PhysRevB.2.4559} {\bibfield  {journal} {\bibinfo  {journal} {Phys. Rev. B.}\ }\textbf {\bibinfo {volume} {2}},\ \bibinfo {pages} {4559} (\bibinfo {year} {1970})}\BibitemShut {NoStop}%
\bibitem [{\citenamefont {Xiao}\ \emph {et~al.}(2010)\citenamefont {Xiao}, \citenamefont {Chang},\ and\ \citenamefont {Niu}}]{xiao2010berry}%
  \BibitemOpen
  \bibfield  {author} {\bibinfo {author} {\bibfnamefont {D.}~\bibnamefont {Xiao}}, \bibinfo {author} {\bibfnamefont {M.-C.}\ \bibnamefont {Chang}},\ and\ \bibinfo {author} {\bibfnamefont {Q.}~\bibnamefont {Niu}},\ }\bibfield  {title} {\bibinfo {title} {Berry {P}hase {E}ffects on {E}lectronic {P}roperties},\ }\href {https://arxiv.org/abs/0907.2021} {\bibfield  {journal} {\bibinfo  {journal} {Rev. Mod. Phys.}\ }\textbf {\bibinfo {volume} {82}},\ \bibinfo {pages} {1959} (\bibinfo {year} {2010})}\BibitemShut {NoStop}%
\bibitem [{\citenamefont {Tian}\ \emph {et~al.}(2009)\citenamefont {Tian}, \citenamefont {Ye},\ and\ \citenamefont {Jin}}]{tian2009proper}%
  \BibitemOpen
  \bibfield  {author} {\bibinfo {author} {\bibfnamefont {Y.}~\bibnamefont {Tian}}, \bibinfo {author} {\bibfnamefont {L.}~\bibnamefont {Ye}},\ and\ \bibinfo {author} {\bibfnamefont {X.}~\bibnamefont {Jin}},\ }\bibfield  {title} {\bibinfo {title} {Proper scaling of the anomalous {H}all effect},\ }\href {https://journals.aps.org/prl/abstract/10.1103/PhysRevLett.103.087206} {\bibfield  {journal} {\bibinfo  {journal} {Phys. Rev. Lett.}\ }\textbf {\bibinfo {volume} {103}},\ \bibinfo {pages} {087206} (\bibinfo {year} {2009})}\BibitemShut {NoStop}%
\bibitem [{\citenamefont {Fang}\ \emph {et~al.}(2003)\citenamefont {Fang} \emph {et~al.}}]{fang2003anomalous}%
  \BibitemOpen
  \bibfield  {author} {\bibinfo {author} {\bibfnamefont {Z.}~\bibnamefont {Fang}} \emph {et~al.},\ }\bibfield  {title} {\bibinfo {title} {The {A}nomalous {H}all {E}ffect and {M}agnetic {M}onopoles in {M}omentum {S}pace},\ }\href {https://www.science.org/doi/full/10.1126/science.1089408} {\bibfield  {journal} {\bibinfo  {journal} {Science}\ }\textbf {\bibinfo {volume} {302}},\ \bibinfo {pages} {92} (\bibinfo {year} {2003})}\BibitemShut {NoStop}%
\bibitem [{\citenamefont {Haldane}(2004)}]{haldane2004berry}%
  \BibitemOpen
  \bibfield  {author} {\bibinfo {author} {\bibfnamefont {F.}~\bibnamefont {Haldane}},\ }\bibfield  {title} {\bibinfo {title} {Berry {C}urvature on the {F}ermi {S}urface: {A}nomalous {H}all {E}ffect as a {T}opological {F}ermi-{L}iquid {P}roperty},\ }\href {https://journals.aps.org/prl/abstract/10.1103/PhysRevLett.93.206602} {\bibfield  {journal} {\bibinfo  {journal} {Phys. Rev. Lett.}\ }\textbf {\bibinfo {volume} {93}},\ \bibinfo {pages} {206602} (\bibinfo {year} {2004})}\BibitemShut {NoStop}%
\bibitem [{\citenamefont {Had{\v{z}}i{\'c}-Leroux}\ \emph {et~al.}(1986)\citenamefont {Had{\v{z}}i{\'c}-Leroux} \emph {et~al.}}]{hadvzic1986hall}%
  \BibitemOpen
  \bibfield  {author} {\bibinfo {author} {\bibfnamefont {M.}~\bibnamefont {Had{\v{z}}i{\'c}-Leroux}} \emph {et~al.},\ }\bibfield  {title} {\bibinfo {title} {Hall {E}ffect in {H}eavy-{F}ermion {S}ystems: {UPt$_{3}$}, {UAl$_{2}$}, {CeAl$_{3}$}, {CeRu$_{2}$Si$_{2}$}},\ }\href {https://iopscience.iop.org/article/10.1209/0295-5075/1/11/006/meta} {\bibfield  {journal} {\bibinfo  {journal} {Europhys. Lett.}\ }\textbf {\bibinfo {volume} {1}},\ \bibinfo {pages} {579} (\bibinfo {year} {1986})}\BibitemShut {NoStop}%
\bibitem [{\citenamefont {Bogach}\ \emph {et~al.}(2003)\citenamefont {Bogach} \emph {et~al.}}]{bogach2003hall}%
  \BibitemOpen
  \bibfield  {author} {\bibinfo {author} {\bibfnamefont {A.}~\bibnamefont {Bogach}} \emph {et~al.},\ }\bibfield  {title} {\bibinfo {title} {Hall effect anomalies in kondo-lattice {CeAl$_{2}$}},\ }\href {https://www.sciencedirect.com/science/article/pii/S0921452602023827} {\bibfield  {journal} {\bibinfo  {journal} {Phys. B. Condens. Matter.}\ }\textbf {\bibinfo {volume} {329}},\ \bibinfo {pages} {541} (\bibinfo {year} {2003})}\BibitemShut {NoStop}%
\bibitem [{\citenamefont {Asaba}\ \emph {et~al.}(2021)\citenamefont {Asaba} \emph {et~al.}}]{asaba2021colossal}%
  \BibitemOpen
  \bibfield  {author} {\bibinfo {author} {\bibfnamefont {T.}~\bibnamefont {Asaba}} \emph {et~al.},\ }\bibfield  {title} {\bibinfo {title} {Colossal anomalous {N}ernst effect in a correlated noncentrosymmetric kagome ferromagnet},\ }\href {https://www.science.org/doi/full/10.1126/sciadv.abf1467} {\bibfield  {journal} {\bibinfo  {journal} {Sci. Adv.}\ }\textbf {\bibinfo {volume} {7}},\ \bibinfo {pages} {eabf1467} (\bibinfo {year} {2021})}\BibitemShut {NoStop}%
\bibitem [{\citenamefont {Blaha}\ \emph {et~al.}(2001)\citenamefont {Blaha} \emph {et~al.}}]{blaha2001wien2k}%
  \BibitemOpen
  \bibfield  {author} {\bibinfo {author} {\bibfnamefont {P.}~\bibnamefont {Blaha}} \emph {et~al.},\ }\bibfield  {title} {\bibinfo {title} {Wien2k},\ }\href {https://susi.theochem.tuwien.ac.at/reg_user/textbooks/usersguide.pdf} {\bibfield  {journal} {\bibinfo  {journal} {An augmented plane wave+ local orbitals program for calculating crystal properties}\ }\textbf {\bibinfo {volume} {60}},\ \bibinfo {pages} {155} (\bibinfo {year} {2001})}\BibitemShut {NoStop}%
\bibitem [{\citenamefont {Kresse}\ and\ \citenamefont {Hafner}(1993)}]{kresse1993ab}%
  \BibitemOpen
  \bibfield  {author} {\bibinfo {author} {\bibfnamefont {G.}~\bibnamefont {Kresse}}\ and\ \bibinfo {author} {\bibfnamefont {J.}~\bibnamefont {Hafner}},\ }\bibfield  {title} {\bibinfo {title} {Ab $initio$ molecular dynamics for open-shell transition metals},\ }\href {https://journals.aps.org/prb/abstract/10.1103/PhysRevB.48.13115} {\bibfield  {journal} {\bibinfo  {journal} {Phys. Rev. B.}\ }\textbf {\bibinfo {volume} {48}},\ \bibinfo {pages} {13115} (\bibinfo {year} {1993})}\BibitemShut {NoStop}%
\bibitem [{\citenamefont {Kresse}\ and\ \citenamefont {Joubert}(1999)}]{kresse1999ultrasoft}%
  \BibitemOpen
  \bibfield  {author} {\bibinfo {author} {\bibfnamefont {G.}~\bibnamefont {Kresse}}\ and\ \bibinfo {author} {\bibfnamefont {D.}~\bibnamefont {Joubert}},\ }\bibfield  {title} {\bibinfo {title} {From ultrasoft pseudopotentials to the projector augmented-wave method},\ }\href {https://journals.aps.org/prb/abstract/10.1103/PhysRevB.59.1758} {\bibfield  {journal} {\bibinfo  {journal} {Phys. Rev. B.}\ }\textbf {\bibinfo {volume} {59}},\ \bibinfo {pages} {1758} (\bibinfo {year} {1999})}\BibitemShut {NoStop}%
\bibitem [{\citenamefont {Perdew}\ \emph {et~al.}(1996)\citenamefont {Perdew}, \citenamefont {Burke},\ and\ \citenamefont {Ernzerhof}}]{perdew1996generalized}%
  \BibitemOpen
  \bibfield  {author} {\bibinfo {author} {\bibfnamefont {J.~P.}\ \bibnamefont {Perdew}}, \bibinfo {author} {\bibfnamefont {K.}~\bibnamefont {Burke}},\ and\ \bibinfo {author} {\bibfnamefont {M.}~\bibnamefont {Ernzerhof}},\ }\bibfield  {title} {\bibinfo {title} {Generalized gradient approximation made simple},\ }\href {https://journals.aps.org/prl/abstract/10.1103/PhysRevLett.77.3865} {\bibfield  {journal} {\bibinfo  {journal} {Phys. Rev. Lett.}\ }\textbf {\bibinfo {volume} {77}},\ \bibinfo {pages} {3865} (\bibinfo {year} {1996})}\BibitemShut {NoStop}%
\bibitem [{\citenamefont {Kune{\v{s}}}\ \emph {et~al.}(2010)\citenamefont {Kune{\v{s}}} \emph {et~al.}}]{kunevs2010wien2wannier}%
  \BibitemOpen
  \bibfield  {author} {\bibinfo {author} {\bibfnamefont {J.}~\bibnamefont {Kune{\v{s}}}} \emph {et~al.},\ }\bibfield  {title} {\bibinfo {title} {Wien2wannier: From linearized augmented plane waves to maximally localized {W}annier functions},\ }\href {https://www.sciencedirect.com/science/article/pii/S0010465510002948} {\bibfield  {journal} {\bibinfo  {journal} {Comput. Phys. Commun.}\ }\textbf {\bibinfo {volume} {181}},\ \bibinfo {pages} {1888} (\bibinfo {year} {2010})}\BibitemShut {NoStop}%
\bibitem [{\citenamefont {Pizzi}\ \emph {et~al.}(2020)\citenamefont {Pizzi} \emph {et~al.}}]{pizzi2020wannier90}%
  \BibitemOpen
  \bibfield  {author} {\bibinfo {author} {\bibfnamefont {G.}~\bibnamefont {Pizzi}} \emph {et~al.},\ }\bibfield  {title} {\bibinfo {title} {Wannier90 as a community code: new features and applications},\ }\href {https://iopscience.iop.org/article/10.1088/1361-648X/ab51ff/meta} {\bibfield  {journal} {\bibinfo  {journal} {J. Phys. Condens. Matter.}\ }\textbf {\bibinfo {volume} {32}},\ \bibinfo {pages} {165902} (\bibinfo {year} {2020})}\BibitemShut {NoStop}%
\bibitem [{\citenamefont {Wu}\ \emph {et~al.}(2018)\citenamefont {Wu} \emph {et~al.}}]{wu2018wanniertools}%
  \BibitemOpen
  \bibfield  {author} {\bibinfo {author} {\bibfnamefont {Q.}~\bibnamefont {Wu}} \emph {et~al.},\ }\bibfield  {title} {\bibinfo {title} {Wannier{T}ools: {A}n open-source software package for novel topological materials},\ }\href {https://www.sciencedirect.com/science/article/pii/S0010465517303442} {\bibfield  {journal} {\bibinfo  {journal} {Comput. Phys. Commun.}\ }\textbf {\bibinfo {volume} {224}},\ \bibinfo {pages} {405} (\bibinfo {year} {2018})}\BibitemShut {NoStop}%
\bibitem [{\citenamefont {Kawakami}\ and\ \citenamefont {Okiji}(1986)}]{kawakami1986magnetoresistance}%
  \BibitemOpen
  \bibfield  {author} {\bibinfo {author} {\bibfnamefont {N.}~\bibnamefont {Kawakami}}\ and\ \bibinfo {author} {\bibfnamefont {A.}~\bibnamefont {Okiji}},\ }\href {https://www.jstage.jst.go.jp/article/jpsj1946/55/7/55_7_2114/_article/-char/ja/} {\bibfield  {journal} {\bibinfo  {journal} {J. Phys. Soc. Jpn.}\ }\textbf {\bibinfo {volume} {55}},\ \bibinfo {pages} {2114} (\bibinfo {year} {1986})}\BibitemShut {NoStop}%
\bibitem [{\citenamefont {Parihari}\ and\ \citenamefont {Vidhyadhiraja}(2009)}]{parihari2009magnetoresistance}%
  \BibitemOpen
  \bibfield  {author} {\bibinfo {author} {\bibfnamefont {D.}~\bibnamefont {Parihari}}\ and\ \bibinfo {author} {\bibfnamefont {N.}~\bibnamefont {Vidhyadhiraja}},\ }\href {https://iopscience.iop.org/article/10.1088/0953-8984/21/40/405602/meta} {\bibfield  {journal} {\bibinfo  {journal} {J. Phys. Condens. Matter}\ }\textbf {\bibinfo {volume} {21}},\ \bibinfo {pages} {405602} (\bibinfo {year} {2009})}\BibitemShut {NoStop}%
\bibitem [{\citenamefont {Cornelius}\ \emph {et~al.}(2000)\citenamefont {Cornelius}, \citenamefont {Arko}, \citenamefont {Sarrao}, \citenamefont {Hundley},\ and\ \citenamefont {Fisk}}]{cornelius2000anisotropic}%
  \BibitemOpen
  \bibfield  {author} {\bibinfo {author} {\bibfnamefont {A.}~\bibnamefont {Cornelius}}, \bibinfo {author} {\bibfnamefont {A.}~\bibnamefont {Arko}}, \bibinfo {author} {\bibfnamefont {J.}~\bibnamefont {Sarrao}}, \bibinfo {author} {\bibfnamefont {M.}~\bibnamefont {Hundley}}, \ and\ \bibinfo {author} {\bibfnamefont {Z.}~\bibnamefont {Fisk}},\ }\href {https://journals.aps.org/prb/abstract/10.1103/PhysRevB.62.14181} {\bibfield  {journal} {\bibinfo  {journal} {Phys. Rev. B}\ }\textbf {\bibinfo {volume} {62}},\ \bibinfo {pages} {14181} (\bibinfo {year} {2000})}\BibitemShut {NoStop}%
\bibitem [{\citenamefont {Rauchschwalbe}\ \emph {et~al.}(1987)\citenamefont {Rauchschwalbe}, \citenamefont {Steglich}, \citenamefont {De~Visser},\ and\ \citenamefont {Franse}}]{rauchschwalbe1987magnetoresistance}%
  \BibitemOpen
  \bibfield  {author} {\bibinfo {author} {\bibfnamefont {U.}~\bibnamefont {Rauchschwalbe}}, \bibinfo {author} {\bibfnamefont {F.}~\bibnamefont {Steglich}}, \bibinfo {author} {\bibfnamefont {A.}~\bibnamefont {De~Visser}}, \ and\ \bibinfo {author} {\bibfnamefont {J.}~\bibnamefont {Franse}},\ }in\ \href {https://www.sciencedirect.com/science/chapter/edited-volume/abs/pii/B9781483229485501064} {\emph {\bibinfo {booktitle} {Anomalous Rare Earths and Actinides}}}\ (\bibinfo  {publisher} {Elsevier},\ \bibinfo {year} {1987})\ pp.\ \bibinfo {pages} {347--350}\BibitemShut {NoStop}%
\bibitem [{\citenamefont {Po}\ \emph {et~al.}(2017)\citenamefont {Po}, \citenamefont {Vishwanath},\ and\ \citenamefont {Watanabe}}]{po2017symmetry}%
  \BibitemOpen
  \bibfield  {author} {\bibinfo {author} {\bibfnamefont {H.~C.}\ \bibnamefont {Po}}, \bibinfo {author} {\bibfnamefont {A.}~\bibnamefont {Vishwanath}}, \ and\ \bibinfo {author} {\bibfnamefont {H.}~\bibnamefont {Watanabe}},\ }\href {https://www.nature.com/articles/s41467-017-00133-2} {\bibfield  {journal} {\bibinfo  {journal} {Nat. Commun.}\ }\textbf {\bibinfo {volume} {8}},\ \bibinfo {pages} {50} (\bibinfo {year} {2017})}\BibitemShut {NoStop}%
\bibitem [{\citenamefont {Bradlyn}\ \emph {et~al.}(2019)\citenamefont {Bradlyn}, \citenamefont {Wang}, \citenamefont {Cano},\ and\ \citenamefont {Bernevig}}]{bradlyn2019disconnected}%
  \BibitemOpen
  \bibfield  {author} {\bibinfo {author} {\bibfnamefont {B.}~\bibnamefont {Bradlyn}}, \bibinfo {author} {\bibfnamefont {Z.}~\bibnamefont {Wang}}, \bibinfo {author} {\bibfnamefont {J.}~\bibnamefont {Cano}}, \ and\ \bibinfo {author} {\bibfnamefont {B.~A.}\ \bibnamefont {Bernevig}},\ }\href {https://journals.aps.org/prb/abstract/10.1103/PhysRevB.99.045140} {\bibfield  {journal} {\bibinfo  {journal} {Phys. Rev. B}\ }\textbf {\bibinfo {volume} {99}},\ \bibinfo {pages} {045140} (\bibinfo {year} {2019})}\BibitemShut {NoStop}%
\bibitem [{\citenamefont {Lin}\ \emph {et~al.}(2026)\citenamefont {Lin} \emph {et~al.}}]{lin2026chiral}%
  \BibitemOpen
  \bibfield  {author} {\bibinfo {author} {\bibfnamefont {K.-S.}\ \bibnamefont {Lin}} \emph {et~al.},\ }\href {https://arxiv.org/pdf/2602.22185} {\bibfield  {journal} {\bibinfo  {journal} {arXiv preprint arXiv:2602.22185}\ } (\bibinfo {year} {2026})}\BibitemShut {NoStop}%

 
\end{thebibliography}
\end{document}